\documentclass[reprint,amsmath,amssymb,pra]{revtex4-2}

\usepackage{graphicx}
\usepackage{dcolumn}
\usepackage{bm}
\usepackage{times}
\usepackage{graphicx}
\usepackage{amsmath}
\usepackage{amsfonts}
\usepackage{mathtools}
\usepackage{setspace}
\usepackage{braket}
\usepackage[table]{xcolor}
\usepackage{makecell}
\usepackage{physics}
\begin{document}

\preprint{APS/123-QED}

\title{Markovianity and non-Markovianity of Particle Bath with Dirac Dispersion Relation}

\author{Takano Taira$^{1~2}$, Naomichi Hatano$^{2}$ and Akinori Nishino$^{2~3}$}

\affiliation{$^1$Department of Physics, Kyushu University, 744 Motooka, Nishi-Ku, Fukuoka 819-0395, Japan \\
$^2$Institute of Industrial Science, The University of Tokyo,5-1-5 Kashiwanoha, Kashiwa, Chiba 277-8574, Japan\\
$^3$Faculty of Engineering, Kanagawa University, 3-27-1 Rokkakubashi, Kanagawa-ku,
Yokohama, Kanagawa, 221-8686, Japan\\
E-mail: taira904@iis.u-tokyo.ac.jp
}

\date{\today}

\begin{abstract}
The dynamics of a two-level system coupled to a particle bath with the Dirac dispersion relation is studied. We analytically show that closing the Dirac gap results in a transition of the survival probability of the two-level system from non-exponential to exponential decay in the long-time regime, while the short-time regime remains exponential. The exact time-evolving state is also calculated. With the Dirac gap closing smoothly, the time-evolving state converges to a time-evolving resonant state, which is normalizable due to causality. We numerically show that introducing a finite cutoff to the Dirac dispersion relation leads to a transition from exponential to non-exponential decay both in short- and long-time regimes, with the time-evolving resonant state resolved to a time-evolving state. Furthermore, we propose several experimental setups that act as a particle bath with the Dirac dispersion relation. We give a detailed calculation for one of them, namely an optical array in the Su–Schrieffer–Heeger configuration. In this case, we show that our theoretical results can be observed experimentally with realistic parameters in an existing experimental setup of an optical waveguide array.

\end{abstract}

\maketitle

\section{Introduction}
%History of the non-exponential decay  
The exponential decay of an unstable system is ubiquitous in natural science. In a quantum system, one of the earliest discoveries of exponential decay is the radioactive decay of an alpha particle, discovered in 1899 by Rutherford~\cite{rutherford1902xxxiii}, and later a theoretical model was proposed by Gamow~\cite{gamow1928quantentheorie}. This was probably the first theoretical analysis of open quantum systems, in which a small quantum system is connected to a large outer space called the environment. 

In Gamow's theory, the decay was assumed to be purely exponential. In fact, Khalfin~\cite{Khalfin1958} showed theoretically in 1958 that the decay profile would deviate from the exponential decay to a power-law decay after a sufficiently long time, if the energy spectrum of the Hamiltonian of the environment is semi-bounded. This power-law decay was experimentally observed much later in 2006 by Rothe \textit{et al}.~\cite{Rothe2006}, as a luminescence decay of an organic material. Under the same condition for the spectrum of the environment, Chiu, Sudarshan, and Misra~\cite{Chiu1977} showed theoretically in 1977 a deviation from the exponential decay in a short-time regime, which led to the prediction of the quantum Zeno effect~\cite{Misra1977}. The Zeno effect was also experimentally confirmed in 1990 by Itano \textit{et al}.~\cite{Itano1990}.

%Under what condition does non-exp appear? 
As stated above, the mechanism behind the non-exponential decay, both in the short- and long-time regimes, is closely related to the semi-bounded spectrum of the Hamiltonian of the environment. More precisely, rigorous analyses show that a non-equilibrium quantum state, called a quasi-stationary state, when connected to the environment with an energy spectrum bounded from below, decays non-exponentially in sufficiently short- and long-time regimes~\cite{Khalfin1958,Chiu1977}. This indicates that every physical system, which naturally assumes that the spectrum is bounded from below, should exhibit non-exponential decay in short- and long-time regimes. A simple example of a quasi-stationary state is an excited state in a qubit, attached to an environment in a vacuum state. The dynamics is governed by an open quantum system in which an unstable system (a qubit) is coupled to an environment.

\begin{figure}[t]
    \centering
    \includegraphics[width=0.4\textwidth]{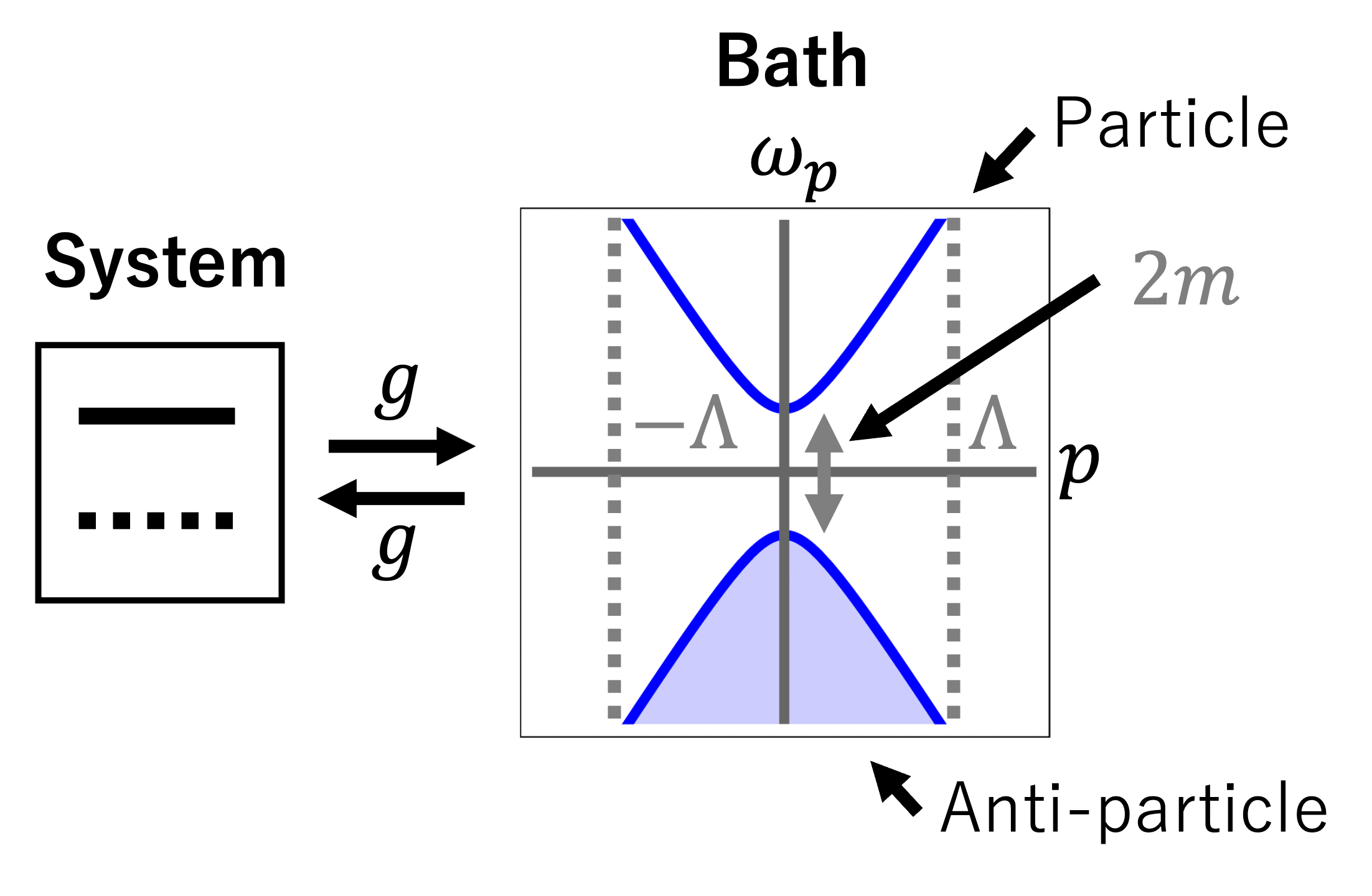}
    \caption{A schematic of our main model. The two-level system is coupled with coupling strength $g$ to the particle bath with the Dirac dispersion relation $\omega_p = \pm\sqrt{p^2 + m^2}$ with a Dirac gap $2m$ and a momentum cutoff $\Lambda$. The plus and minus branches of the dispersion correspond to the particle and the anti-particle.
}
    \label{fig:schematics}
\end{figure}

%Given above paragraph, why can't we see non-exp in experiment?  
The difficulty of observing non-exponential decay in  experiments can be attributed to its time scale. The short-time dynamics is characterized by the Zeno time, given by the inverse square root of the variance of the Hamiltonian with respect to the initial quantum state. Such a quantity is large when one considers the entire Hamiltonian, including the surrounding environment, and hence the time scale is almost zero. There is no universal time scale for the long-time regime, but there have been several works in which it is often associated with the low-energy structure of the spectrum, such as the gap between the bound state energy and the continuum of the spectrum~\cite{garmon2013amplification,Kofman1994,Redondo-Yuste2021,Gonzalez2018}.

%Novelty of our work and why it's different from previous works  
Previous studies of non-exponential decay have primarily focused on either the short-time or long-time regime, but not both simultaneously. Furthermore, they typically assume a semi-bounded spectrum with a spectral gap, treating these features as fixed properties of the environment rather than controllable parameters. In the present work, we study an interplay between the spectral structure of the environment and the short- and long-time dynamics. We propose a two-level system coupled to a Dirac environment, schematically shown in Fig.~\ref{fig:schematics}, in which we control the gap and the semi-boundedness of the spectrum by adjusting the Dirac gap $2m$ and the cutoff $\Lambda$. Our key findings are summarized in Table~\ref{tab:results}. 

Furthermore, we analyze the time evolution of a time-evolving state that leaks from the system into the environment and observe that, as the Dirac gap closes and the cutoff is taken to infinity, the time-evolving state smoothly converges into a normalizable time-evolving resonant state~\cite{HatanoNishino,petrosky2001space}. The time-evolving resonant state is a wave function that increases exponentially in space but has only a finite support due to the causality constraint. Its profile coincides with the resonant eigenstate within the causal region $\left|x\right|<c t$, where $c$ is the speed of light, but drops to zero outside~\cite{HatanoNishino,petrosky2001space}. Note that the resonant eigenstate itself is an eigenstate of a Hamiltonian under the Siegert boundary condition with a complex eigenvalue~\cite{gamow1928quantentheorie,siegert1939derivation,landau2013quantum},  which is non-normalizable due to its lack of a finite spatial support, whereas the time-evolving resonant state is a solution to the initial-value problem of the Schrödinger equation.

\begin{table}[!htb]
\begin{minipage}{.5\textwidth}
  \text{Short-time decay profile of survival prob.}
  \centering 
    \begin{tabular}{ccc}\hline
         & $m=0$ & $m\not= 0$\\\hline
        $\Lambda < \infty$&\cellcolor{lightgray}Quadratic &\cellcolor{lightgray}Quadratic \\\hline
        $\Lambda \rightarrow \infty$&Exp.  &Exp. \\\hline
    \end{tabular}
\end{minipage}\\[0.5cm]
\begin{minipage}{.5\textwidth}
  \centering
    \text{Long-time decay profile of survival prob.}
    \begin{tabular}{ccc}\hline
         & $m=0$ & $m\not= 0$\\\hline
        $\Lambda < \infty$&\makecell{\cellcolor{lightgray}$\mathcal{O}(t^{-1})$ decay\\ \cellcolor{lightgray}to bound state}&\makecell{\cellcolor{lightgray}Power-law decay \\ \cellcolor{lightgray}to bound state} \\\hline
        $\Lambda \rightarrow \infty$&\makecell{Exp. decay \\ to zero }&\makecell{\cellcolor{lightgray}$\mathcal{O}(t^{-3/2})$ decay to \\ \cellcolor{lightgray}bound state}\\\hline
    \end{tabular}
\end{minipage} 
\caption{Two tables showing the decay profiles of the survival probability in the short- and long-time regimes. The two parameters $m$ and $\Lambda$ of our model are schematically shown in Fig.~\ref{fig:schematics}.}
\label{tab:results}
\end{table}

In Section \ref{Sec:model}, we introduce our two-level system coupled to a Dirac environment, exactly derive a non-Markovian dynamical equation and solve it, finding the wave function. We will also comment on the connection between the non-exponential decay and the Markovianity of the dynamics. In Section \ref{Sec:Spectural_structure_and_Markovianity}~A--C, we analytically investigate the exact non-Markovian dynamical equation in the three cases: $m=0, \Lambda\rightarrow\infty$; $m\not= 0, \Lambda\rightarrow \infty$; and $m=0, \Lambda< \infty$. In Section \ref{sec: full analysis}, we numerically investigate the most general case: $m>0, \Lambda<\infty$. Section \ref{sec: experiment} presents three possible experimental setups, with a special focus on the optical waveguide array. A summary is given in Section \ref{Sec: conclusion}.

\section{Survival probability and wave function of a two-level system coupled to a Dirac environment}\label{Sec:model}

\subsection{Model}
In order to construct a model that incorporates the gap and the cutoff structure, we consider an extension of the Friedrichs–Lee model~\cite{lee1954some,friedrichs1948perturbation} but with the Dirac Hamiltonian as the environment Hamiltonian:
\begin{align}\label{Eq:Our_model}
    H_{\text{full}} = \omega_0 \sigma_+ \sigma_- +  H_D + \left(\sigma_+ \otimes B +\sigma_- \otimes  B^\dagger\right).
\end{align}
The first term of the full Hamiltonian (\ref{Eq:Our_model}) represents a two-level system with the eigenstates $\{\ket{\mathrm{g}}_S,\ket{\mathrm{e}}_S \}$, corresponding to the ground and excited states, respectively. We take the level difference $\omega_0$ to be a positive constant. The operators $\sigma_{+} := (\sigma_x - i \sigma_y ) /2$ and $\sigma_- := (\sigma_x + i \sigma_y ) /2$ are the creation, $\ket{\mathrm{e}}_S= \sigma_+ \ket{\mathrm{g}}_S$, and annihilation $\ket{\mathrm{g}}_S= \sigma_- \ket{\mathrm{e}}_S$ operators of the two-level system, which satisfy  fermionic anticommutation relations:
\begin{align}
    \{\sigma_-, \sigma_+\} = \{\sigma_+, \sigma_-\} = 1, \quad \{\sigma_-, \sigma_-\} = \{\sigma_+, \sigma_+\} = 0,
\end{align}
where curly brackets indicate the anticommutator.

The second term of the full Hamiltonian (\ref{Eq:Our_model}) denotes a particle bath of the Dirac Hamiltonian, which we refer to as the \textit{environment}:
\begin{align}\label{Eq:Defining_Dirac_Hamiltonian_and_vacuum}
    H_D = \int_{-\Lambda}^{\Lambda} dp~ \omega_p \left(b^\dagger_p b_p - c_p c_p^\dagger\right),
\end{align}
where $\Lambda>0$ is the cutoff in momentum space, and the Dirac dispersion relation $\omega_p$ corresponds to either massive or massless Dirac particles:
\begin{align}
\text{massive:} \quad & \omega_p = \sqrt{p^2 + m^2}, \label{Eq:massive_dispersion} \\
\text{massless:} \quad & \omega_p = |p|, \label{Eq:massless_dispersion}
\end{align} 
where $m$ is the Dirac mass that generates the Dirac gap $2m$ in the spectrum. The operators $b^\dagger_p, b_p$ and $c^\dagger_p, c_p$ are fermionic creation and annihilation operators of particles and anti-particles, respectively, satisfying anticommutation relations:
\begin{align}
    &\{b_p, b^\dagger_k\} = \{c_p, c^\dagger_k\} = \delta(p-k).
\end{align}
All other anticommutation relations are zero. The vacuum state is defined as $\ket{\text{vac}}_E := \prod_p c_p^\dagger \ket{0}$. This Hamiltonian can be understood as an effective theory near the Fermi level in condensed-matter physics, where operators $b^\dagger_p $ and $ b_p $ correspond to particles above the Fermi level, while $c^\dagger_p $ and $ c_p $ correspond to holes below it.

Finally, the last two terms of the full Hamiltonian (\ref{Eq:Our_model}) represent the interaction between the two-level system and the environment. The third term, $\sigma_+ \otimes B$, where $B := \int_{-\Lambda}^\Lambda dp~ g_p B_p$ with $B_p := b_p + c_p^\dagger$, excites the two-level system while simultaneously either destroying a particle via $b_p$ or creating an anti-particle via $c_p^\dagger$ in the environment. The operators  $\sigma_+ \otimes B$ and $\sigma_- \otimes B^\dagger$ commute with the number operator $N=\sigma_+ \sigma_- + \int_{-\Lambda}^{\Lambda} dp (b_p^\dagger b_p +  c_p c^\dagger_p )$, which would ensure that the overall particle number is preserved. The coupling parameter $g_p$ captures the details of the interaction. Throughout this paper, we consider the case of constant coupling, $g_p = g$. The Fourier transform of our model to real space \(x\) can be understood as a single qubit attached to the origin of a one-dimensional space with a dispersion relation given by the Dirac dispersion \(\omega_p\).

An alternative environmental Hamiltonian can be considered when the momentum range \([-\Lambda, \Lambda]\) is identified with the Brillouin zone \([-\pi, \pi]\), and the dispersion relation \(\omega_p\) is taken to be a cosine band. This leads to the Su–Schrieffer–Heeger (SSH) model~\cite{su1979solitons}. In this context, the positive-energy band $\omega_p$ corresponds to the conduction band, while the negative-energy band $-\omega_p$ corresponds to the valence band, with the Dirac mass gap given by the difference in the hopping amplitudes of the SSH model. We will discuss this case in detail in Sec.~\ref{Sec: optical experiment} of the present work.

\subsection{Exact non-Markovian dynamical equation for the spontaneous emission of the two-level system}\label{Sec: Master equation}

In this section, we derive the exact dynamical equation governing the dynamics of the two-level system and obtain its formal solutions. We also introduce two key quantities: the survival probability, which will be used to analyze the non-exponential decay of the system, and the wave function in the Dirac environment, which allows us to investigate how the time evolution of the state depends on the parameter choices of our model. The results in this section provide the basis for the calculations presented in  Sec.~\ref{Sec:Spectural_structure_and_Markovianity}.

Let us derive the exact non-Markovian dynamical equation associated with our model. For brevity, we use the abbreviation $\ket{\mathrm{g}~\text{vac}} := \ket{\mathrm{g}}_S \otimes \ket{\text{vac}}_E$ for the composite state. The excited state is defined as $\ket{\mathrm{e}~\text{vac}} := \sigma_+\ket{\mathrm{g}~\text{vac}}$. Following the procedure outlined in Refs.~\cite{garraway1997nonperturbative,breuer2002theory}, we assume that the initial state contains the excited state of the two-level system without any particle in the environment:
\begin{align}\label{Eq:initial_condition}
    \ket{\phi(0)} &:= \ket{\mathrm{e}~\text{vac}},
\end{align}
and solve the problem of spontaneous emission. The state at an arbitrary time $t$ can be expressed using the following Ansatz:
\begin{align}\label{Eq: Ansatz}
    \ket{\phi(t)} &:= \phi(t) \ket{\mathrm{e}~\text{vac}} \nonumber\\
    &+ \int_{-\Lambda}^{\Lambda} dp~ \phi_p(t) b_p^\dagger \ket{\mathrm{g}~\text{vac}} + \int_{-\Lambda}^{\Lambda} dp~ \psi_p(t) c_p \ket{\mathrm{g}~\text{vac}},
\end{align}
where the expansion coefficients $\phi(t)$, $\phi_p(t)$, and $\psi_p(t)$ satisfy the normalization condition
\begin{align}\label{Eq:Unity_relation}
    \left|\phi(t)\right|^2 + \int_{-\Lambda}^{\Lambda} dp~ \left|\phi_p(t)\right|^2 + \int_{-\Lambda}^{\Lambda} dp~ \left|\psi_p(t)\right|^2 = 1
\end{align}
and the initial conditions are set to $\phi(0)=1$ and $\phi_p(0)=\psi_p(0)=0$.

By substituting the Ansatz (\ref{Eq: Ansatz}) into the Schrödinger equation for the full system, $i \hbar \partial_t \ket{\phi (t)} = H_{\text{full}} \ket{\phi (t)}$, we obtain the following coupled first-order equations for the coefficients $\phi (t)$, $\phi_p (t)$, and $\psi_p (t)$:
\begin{align}\label{Eq:interaction_picture_Schrodinger_equation}
     i\hbar \frac{d\phi(t)}{dt} &= \omega_0 \phi(t) + g \int_{-\Lambda}^{\Lambda} dp ~ (\phi_p(t) + \psi_p(t)),\\
     i\hbar \frac{d\phi_p(t)}{dt} &= \omega_p \phi_p(t) + g \phi(t),\\
     i\hbar \frac{d\psi_p(t)}{dt} &= -\omega_p \psi_p(t) + g \phi(t).
\end{align}
To eliminate the first term on the right-hand side of each equation, we introduce the new functions $\Phi(t) := e^{i \omega_0 t / \hbar} \phi(t)$, $\Phi_p(t) := e^{i \omega_p t / \hbar} \phi_p(t)$, and $\Psi_p(t) := e^{-i \omega_p t / \hbar} \psi_p(t)$ in the interaction picture. The coupled equations then take the following form:
\begin{align}
    i \hbar \frac{d\Phi(t)}{dt} &= g \int_{-\Lambda}^{\Lambda} dp~ e^{\frac{i}{\hbar} \omega_0 t} \left(\Phi_p(t) e^{-\frac{i}{\hbar} \omega_p t} + \Psi_p(t) e^{\frac{i}{\hbar} \omega_p t} \right), \label{Eq:master_equation_step2}\\
    i \hbar \frac{d\Phi_p(t)}{dt} &= g e^{-\frac{i}{\hbar} (\omega_0 - \omega_p)t} \Phi(t), \\
    i \hbar \frac{d\Psi_p(t)}{dt} &= g e^{-\frac{i}{\hbar} (\omega_0 + \omega_p)t} \Phi(t).
\end{align}

Solving the last two equations, we obtain
\begin{align}\label{Eq. solution to phi p}
    \Phi_p(t) &= \frac{1}{i\hbar} g \int_0^t ds~ e^{-\frac{i}{\hbar} (\omega_0 - \omega_p) s} \Phi(s) + \Phi_p(0), \\
    \Psi_p(t) &= \frac{1}{i\hbar} g \int_0^t ds~ e^{-\frac{i}{\hbar} (\omega_0 + \omega_p) s} \Phi(s) + \Psi_p(0).
\end{align}
By inserting these into Eq.~(\ref{Eq:master_equation_step2}), we find that $\Phi(t)$ obeys the following exact non-Markovian dynamical equation:
\begin{align}\label{Eq:master_equation}
    \frac{d\Phi(t)}{dt} &= \frac{-2g^2}{\hbar^2} \int_0^t ds~ e^{\frac{i}{\hbar} \omega_0 (t - s)} \Phi(s) \int_{-\Lambda}^{\Lambda} dp~ \cos \left( \omega_p \frac{t - s}{\hbar} \right) \nonumber\\
    &\quad - i\frac{g}{\hbar} \int_{-\Lambda}^{\Lambda} dp~ e^{\frac{i}{\hbar} \omega_0 t} \left( e^{-\frac{i}{\hbar} \omega_p t} \Phi_p(0) + e^{\frac{i}{\hbar} \omega_p t} \Psi_p(0) \right).
\end{align}
Under the initial conditions $\Phi_p(0) = \Psi_p(0) = 0$, this simplifies to
\begin{align}
    \frac{d\Phi(t)}{dt} &= \int_0^t ds~ \mathcal{K}(t - s) \Phi(s), \label{Eq:Master equation simple form} 
\end{align}
where $\mathcal{K}$ denotes the memory kernel
\begin{align}
    \mathcal{K}(t) &:= - \frac{2g^2}{\hbar^2} e^{\frac{i}{\hbar} \omega_0 t} \int_{-\Lambda}^{\Lambda} dp~ \cos\left(\omega_p \frac{t}{\hbar}\right).\label{Eq:Memory_kernel}
\end{align}

Using the Laplace transform method on the above exact dynamical equation~(\ref{Eq:Master equation simple form}), as detailed in Appendix~\ref{Appdx: Laplace}, yields the formal solution as an inverse Laplace transform in the form of Bromwich integral:
\begin{align}
    \Phi(t) &= e^{\frac{i}{\hbar} \omega_0 t} \lim_{R \rightarrow \infty} \int_{-iR + \sigma}^{iR + \sigma} \frac{dz}{2\pi i} e^{g^2\frac{z}{\hbar} t} F(z) \Phi(0), \label{Eq:full_solution} \\
    F(z) &:= \left[z + i\frac{\omega_0}{g^2} +  \frac{4z \, \text{arctan}\left( \frac{\Lambda / g^2}{\sqrt{(m/g^2)^2 + z^2}}   \right)}{\sqrt{(m/g^2)^2 + z^2}} \right]^{-1}, \label{Eq:resolvent}
\end{align}
where we have expressed the integrand in terms of the dimensionless parameters \( \omega_0 / g^2 \), \( m / g^2 \) and \(\Lambda / g^2\). The parameter $\sigma$ is an arbitrary real parameter greater than the real part of any poles of the integrand. This solution is employed in Sec.~\ref{Sec: short-time} and Sec.~\ref{sec: short time with finite cut-off}.

%Paragraph: formal solution
The formal solution~(\ref{Eq:full_solution}) is useful for the analysis of the dynamics in the long-time regime. For the short-time regime, it is more convenient to employ the formal solution of Eq.~(\ref{Eq:Master equation simple form}) in terms of the Dyson series:
\begin{align}\label{Eq. Dyson series}
    \Phi(t) = \mathcal{T} \exp\left( \int_0^t ds \int_0^{t - s} du~ \mathcal{K}(u) \right),
\end{align}
where $\mathcal{T} \exp(\dots)$ denotes the time-ordered exponential. This solution is employed in Sec.~\ref{Sec: Bound state} and Sec.~\ref{Sec: short time with finite cut-off}. 

%Paragraph: Key quantities
Finally, let us define the two key quantities of interest. The first is the survival probability of the initial state (\ref{Eq:initial_condition}), given by
\begin{align}\label{Eq: Survival Probability}
    P(t) := \left| \braket{\phi(0) }{ \phi(t)} \right|^2 = \left| \phi(t) \right|^2 = \left| \Phi(t) \right|^2.
\end{align}
This will be used to analyze the appearance of non-exponential decay for different parameter choices. The second key quantity is the wave function of the final state $\ket{\phi(t)}$ in the Dirac environment, in position representation: 
\begin{align}
    \psi(t,x) &:= \braket{\mathrm{g}~x }{ \phi(t)} \nonumber\\
    &= \int_{-\Lambda}^{\Lambda} dp~ \left[e^{-\frac{i}{\hbar} (\omega_p t + px)} \Phi_p(t) + e^{\frac{i}{\hbar} (\omega_p t + px)} \Psi_p(t) \right] \nonumber\\
    &= \frac{1}{i} \sqrt{\frac{2g^2 \pi}{\hbar}} \lim_{R \rightarrow \infty} \int_{-iR + \sigma}^{iR + \sigma} \frac{dz}{2\pi i} e^{\frac{z}{\hbar} t}  F(z/g^2)/g^2 \nonumber\\
    &\quad \times \int_{-\Lambda}^\Lambda dp~ \frac{z \cos\left( \frac{p x}{\hbar} \right) }{z^2 + \omega_p^2}, \label{Eq: Wave function}
\end{align}
where $-i\sqrt{2g^2 \pi / \hbar}$ is the normalization constant. See Appendix~\ref{Appdx: wave function} for a detailed derivation. Here, the state $\ket{\mathrm{g}~x}$ is defined as the Fourier transform of the creation and annihilation operators $b_p^\dagger$ and $c_p$ applied to the vacuum state:
\begin{align}\label{Eq. position representation}
    \ket{\mathrm{g}~x} := \int_{-\Lambda}^\Lambda dp \left( e^{\frac{i}{\hbar} p x} b_p^\dagger + e^{-\frac{i}{\hbar} p x} c_p \right) \ket{\mathrm{g}~\text{vac}}.
\end{align}
The wave function will be used to analyze how the time-evolving state converges to the time-evolving resonant state by controlling the parameters.

\subsection{Connection to the Markovianity}\label{Sec: Connection to the Markovianity}

Before we proceed to the analysis of our model, let us briefly comment on the connection between the decay profile and the Markovianity in our setup.

In an open quantum system, the system’s dynamics is governed by the Nakajima–Zwanzig equation~\cite{Nakajima1958,Zwanzig1960}, which is an integro-differential equation for the system's density matrix \(\rho_S(t)\). Under the Born–Markov approximation~\cite{davies1974markovian,breuer2002theory}, this equation is reduced to the dynamical semigroup form known as the Gorini–Kossakowski–Sudarshan–Lindblad (GKSL) equation~\cite{Gorini1976,Lindblad1976}:
\begin{align}
    \frac{\partial \rho_S}{\partial t} = \mathcal{L} \left[\rho_S(t)\right],
\end{align}
where \(\mathcal{L}\) is the GKSL Liouvillian superoperator, acting on a density operator. Throughout this work, we refer to the dynamics governed by the GKSL equation as the Markovian dynamics, though we note that the Markovianity can be defined in multiple ways depending on the physical context~\cite{Li2018}. In Appendix~\ref{Appdx: CP-divisibility}, we further assess the one-parameter quantum dynamical semigroup property, which we simply call the semigroup property, as another characterization of Markovianity in the present setup. This property is associated with the existence of a time-independent GKSL Liouvillian superoperator $\mathcal{L}$, under which the reduced dynamics is generated as $\exp(t\mathcal{L})$~\cite{Gorini1976,Lindblad1976}. This condition is stronger than the complete-positivity divisibility (CP-divisibility), which is also used to characterize Markovianity. In the present work, we focus on the semigroup property and do not discuss the general CP-divisibility.

Let us consider the initial state \(\ket{\psi(0)} = \ket{\mathrm{e}}_S \otimes \ket{b_0}_E\), where \(\ket{\mathrm{e}}_S \in \mathbb{C}^2\) denotes the excited state of the two-level system and \(\ket{b_0}_E \in \mathcal{H}_E\) is an initial state of the environment. In principle, there is freedom to choose any environment state $\ket{b_0}_E $, but for a single-particle setting with our initial condition, this is fixed to a vacuum state $\ket{\text{vac}}_E\in\mathcal{H}_E$. In this case, it can be shown that if the GKSL Liouvillian superoperator is diagonalizable, the survival probability decomposes into the sum of purely exponential decay terms and exponentially decaying oscillations (see Appendix~\ref{Appdx: Markov}).

If the GKSL Liouvillian superoperator is non-diagonalizable, the survival probability additionally contains exponential decaying terms with polynomial prefactors with the degree \(n\), which is determined by the rank of the Jordan block associated with the non-diagonalizable part of \(\mathcal{L}\) (see Appendix~\ref{Appdx: Markov}).

As a result, non-exponential decay at either short or long times cannot be achieved if we limit ourselves to GKSL dynamics, making it necessary to adopt non-Markovian approaches. Furthermore, we show in Appendix~\ref{Appdx: CP-divisibility} that non-exponential decay violates the semigroup property, which indicates non-Markovianity in the sense of our definition of Markovianity.

In the next section, we analyze the survival probability (\ref{Eq: Survival Probability}) and the wave function (\ref{Eq: Wave function}) in different parameter regimes, focusing on the Dirac gap \(2m\) and the momentum cutoff \(\Lambda\).

\section{Influence of the Dirac Gap and the Spectral Cutoff on Decay Dynamics and the Wave Function}\label{Sec:Spectural_structure_and_Markovianity}

In this section, we investigate how the Dirac gap \(2m\) and the spectral cutoff \(\Lambda\) of the environment Hamiltonian~(\ref{Eq:Defining_Dirac_Hamiltonian_and_vacuum}) influence both the decay profile of the survival probability~(\ref{Eq: Survival Probability}) and the wave function in the position representation~(\ref{Eq: Wave function}). 
The next four subsections are devoted to analyzing the following distinct parameter cases: Sec.~\ref{sec: L infinite and m zero} \((m = 0,\, \Lambda \to \infty)\), Sec.~\ref{Sec:Low_energy_structure_and_Markovianity} \((m \neq 0,\, \Lambda \to \infty)\), Sec.~\ref{Sec:High_energy_structure_and_Markovianity} \((m = 0,\, \Lambda < \infty)\), and Sec.~\ref{sec: full analysis} \((m \neq 0,\, \Lambda < \infty)\).

Two main findings emerge from this analysis. 
First, when the Dirac gap is closed (\(2m = 0\)) and the cutoff is taken to infinity (\(\Lambda \rightarrow \infty\)), the survival probability exhibits a purely exponential decay across all time regimes, in contrast to the findings of Khalfin~\cite{Khalfin1958}, Chiu, Sudarshan and Misra~\cite{Chiu1977}. We also find that the wave function corresponds to a time-evolving resonant state~\cite{HatanoNishino,petrosky2001space}, which is normalizable due to the causality constraint. 
Second, while opening the Dirac gap does not affect the short-time behavior, it causes a deviation from exponential to power-law decay in the long-time regime and dissolution of the time-evolving resonant state. 
On the other hand, introducing a spectral cutoff leads to non-exponential, power-law decay even in the short-time regime. These results are summarized in Table~\ref{tab:results}.

\subsection{Exponential decay: $m=0$ and $\Lambda\rightarrow \infty$}\label{sec: L infinite and m zero}

\subsubsection{Survival probability}

In the case \( m = 0 \) and \( \Lambda \rightarrow \infty \), the survival probability~(\ref{Eq: Survival Probability}) exhibits a purely exponential decay. To see this, we analyze the non-Markovian dynamical equation~(\ref{Eq:Master equation simple form}). 
In the limits $m=0$ and $\Lambda\rightarrow \infty$, the memory kernel $\mathcal{K}$ takes the following form:
\begin{align}
    \mathcal{K}(t) &= - \frac{2 g^2}{\hbar^2} e^{\frac{i}{\hbar}\omega_0 t} \int_{-\infty}^{\infty} dp~ \cos\left(|p| \frac{t}{\hbar}\right) \nonumber\\
    &= - \frac{2 g^2}{\hbar^2} e^{\frac{i}{\hbar}\omega_0 t}  \pi \hbar \delta(t),\label{Eq: memory kernel in Marko limit}
\end{align}
where $\delta (t)$ is Dirac's delta function. Notice that this limit simplifies the non-Markovian dynamical equation~(\ref{Eq:Master equation simple form}) to a time-local linear differential equation. This implies that the dynamics of the system is Markovian, and therefore we refer to this limit as the \textit{Markovian limit}. In an alternative definition based on the semigroup property, this limit also corresponds to Markovianity, as shown in Appendix~\ref{Appdx: CP-divisibility}.

Substituting Eq.~(\ref{Eq: memory kernel in Marko limit}) into Eq.~(\ref{Eq:Master equation simple form}), we obtain the solution for the probability amplitude and the corresponding survival probability as follows:
\begin{align}\label{Eq: Surv. Prob. 1}
    \Phi(t) = e^{- \frac{2g^2 \pi}{\hbar} t}, \quad P(t) = \left|\Phi(t)\right|^2 = \left|\phi(t)\right|^2 = e^{- \frac{4g^2 \pi}{\hbar} t},
\end{align}
which shows a purely exponential decay.

It is worth noting that, according to the works of Khalfin~\cite{Khalfin1958} and Chiu, Sudarshan, and Misra~\cite{Chiu1977}, a purely exponential decay over the short- and long-time regimes implies that the Hamiltonian must be unbounded. However, the converse is not necessarily true: an unbounded Hamiltonian does not guarantee an exponential decay. A counterexample can be found in Ref.~\cite{burgarth2017positive}. In our model, we explicitly observe a purely exponential decay across all timescales, making it an example of the converse of the sufficient condition. In the next subsection \ref{Sec:Low_energy_structure_and_Markovianity}, we show that by introducing the spectral gap \( 2m \neq 0 \), the system exhibits an exponential decay only in the short-time regime while deviating from the exponential decay at long times. This provides an example of a non-exponential decay arising from an unbounded Hamiltonian, thereby serving as a counterexample to the converse statement by Khalfin, Chiu, Sudarshan, and Misra~\cite{Khalfin1958,Chiu1977}.

\subsubsection{Wave function of environmental states}

To obtain the wave function, we take the limit $m\rightarrow 0$ then $\Lambda\rightarrow \infty $ to the formal solution~(\ref{Eq: Wave function}) and find the following:
\begin{align}
    \psi(t,x) &= \frac{1}{i}\sqrt{\frac{2g^2 \pi}{\hbar}} \lim_{R \rightarrow \infty} \int_{-iR + \sigma}^{iR + \sigma} \frac{dz}{2\pi i} e^{\frac{z}{\hbar} t}  \nonumber\\
    &\quad \times \frac{\pi z e^{-\frac{\left|x\right|}{\hbar}\sqrt{z^2}}}{(z+i \omega_0)\sqrt{z^2}+2g^2 \pi z}\nonumber\\
    &= \frac{1}{i}\sqrt{\frac{2g^2 \pi}{\hbar}} \text{Lap}^{-1} \left[ \frac{\pi z e^{-\frac{\left|x\right|}{\hbar}\sqrt{z^2}}}{(z+i \omega_0)\sqrt{z^2}+2g^2 \pi z},\frac{t}{\hbar}\right].
\end{align}
Note that the order of limits does not affect the end result. Finally, evaluating the inverse Laplace transform, we obtain the wave function 
\begin{align}\label{Eq: Exp wave function}
    \psi(t,x) = \frac{1}{i} \sqrt{\frac{2g^2\pi}{\hbar}} e^{-\frac{i}{\hbar} \omega_0 (t - |x|)} e^{- \frac{2g^2 \pi}{\hbar} (t - |x|)}\, \theta(t - |x|),
\end{align}
where \( \theta(t - |x|) \) is the Heaviside step function.

\begin{figure}[t]
    \centering
    \includegraphics[width=0.45\textwidth]{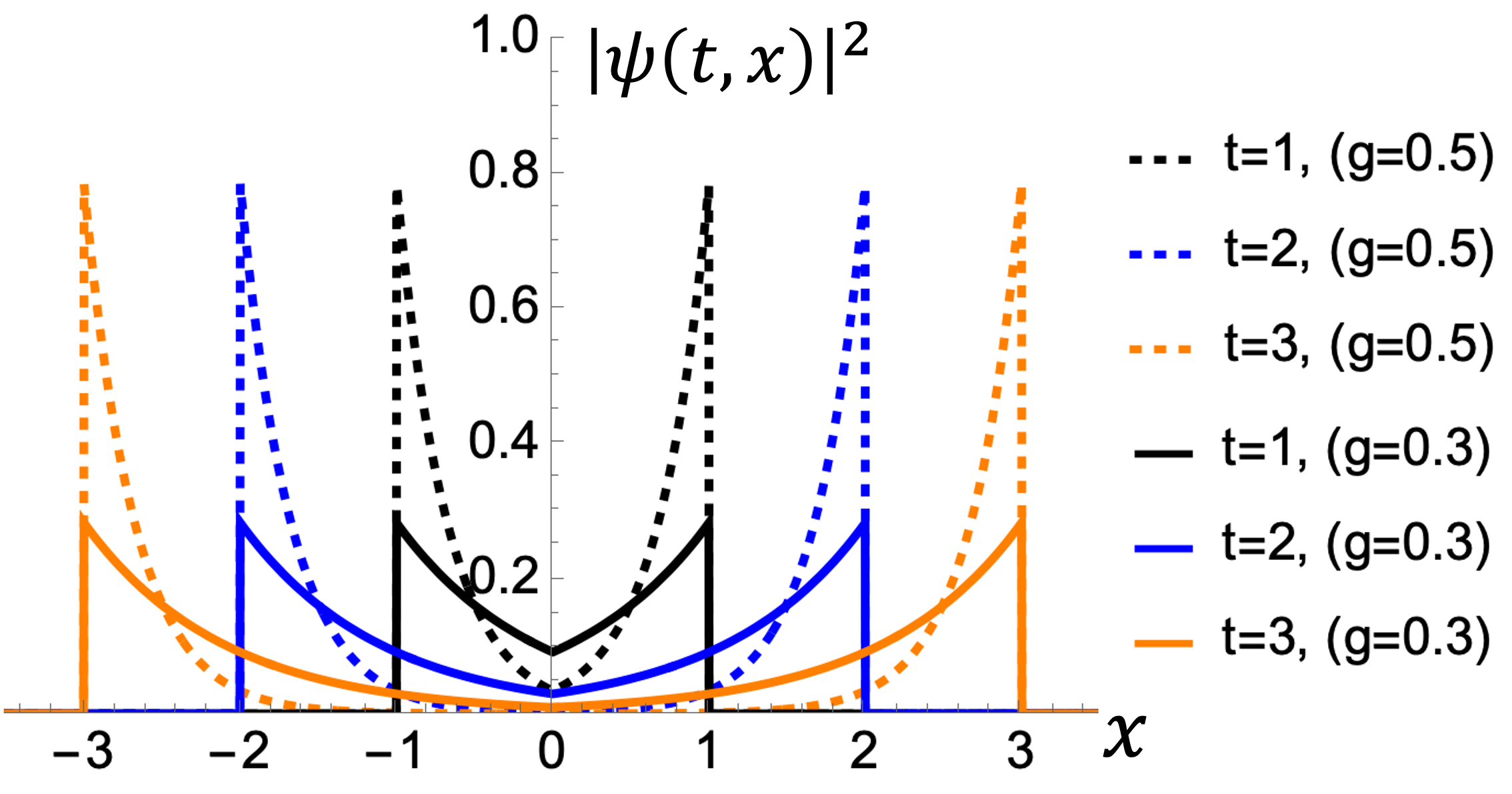}
    \caption{Plot of the existence probability corresponding to the wave function in Eq.~(\ref{Eq: Exp wave function}), with the parameters \( \hbar = 1 \), \( \omega_0 = 1 \).}
    \label{fig:Exp_decay_and_wave}
\end{figure}

The existence probability \( \left|\psi(t,x)\right|^2 \) derived from this wave function is shown in Fig.~\ref{fig:Exp_decay_and_wave}. We observe a finite-peak structure at the points $x=\pm t$, followed by an abrupt drop to zero due to the presence of the step function in Eq.~(\ref{Eq: Exp wave function}). This step function enforces the causality condition, ensuring that the wave function has a support only within the light cone \( t = |x| \). Such a wave function has been referred to as a \textit{time-evolving resonant state} in Ref.~\cite{HatanoNishino} and was previously observed theoretically in Ref.~\cite{petrosky2001space}.

By combining the survival probability of the two-level system in Eq.~(\ref{Eq: Surv. Prob. 1}) with the probability associated with the time-evolving resonant state~(\ref{Eq: Exp wave function}), we find that the total probability is conserved:
\begin{align}
    |\phi(t)|^2 + \int_{-\infty}^{\infty} dx\, |\psi(t,x)|^2 = 1.
\end{align}
This conservation implies that the time-evolving resonant state admits a standard probabilistic interpretation in accordance with the Born rule.

Lastly, let us comment on the normalizability of the standard resonant state. The resonant state is a solution to an eigenvalue problem with a complex eigenvalue, leading to an exponentially increasing wave function in space. Consequently, the issue of non-normalizability naturally emerges. Numerous previous studies have therefore attempted to impose normalizability on the resonant state, as summarized in Ref.~\cite{lalanne2018light}. Recently, an extension of the Hilbert space to the rigged Hilbert space in order to normalize the resonant state was introduced~\cite{de2012rigged}. One of the authors (N.H.) of the present work has also contributed by proving that the resonant state can have a probabilistic interpretation with moving boundaries of the integration domain \cite{hatano2008some,hatano2009probabilistic}. Our result, Eq.~(\ref{Eq: Exp wave function}), similarly exhibits a wave function that increases exponentially in space. Despite this similar spatial behavior, our solution is not a standard resonant state. It is not a solution to the eigenvalue problem, but rather to an initial-value problem, and hence it should instead be regarded as a time-evolving resonant state~\cite{petrosky2001space,HatanoNishino}. In contrast to the standard resonant state, this state automatically has a finite support and therefore requires no additional normalization structure.

\subsection{Low-energy structure: \(m>0\) and \( \Lambda \rightarrow \infty\)}\label{Sec:Low_energy_structure_and_Markovianity}
    %Opening
    The main objective of this subsection is to examine how a finite Dirac gap, \( 2m \neq 0 \), affects the exponential decay and the time-evolving resonant state identified in Sec.~\ref{sec: L infinite and m zero}. 
    To isolate this effect, we keep the limit \( \Lambda \rightarrow \infty \) in the memory kernel~(\ref{Eq:Memory_kernel}) and the wave function~(\ref{Eq: Wave function}). 
    As the expressions for the survival probability and the wave function in this regime are substantially more involved than in the Markovian limit discussed in the previous section~\ref{sec: L infinite and m zero}, we begin by outlining the method used to analyze their behavior, before turning to a detailed asymptotic analysis in the short- and long-time regimes. Furthermore, we will use different approaches for the short- and long-time regimes. In the long-time regime analyzed in Sec.~\ref{Sec: short-time}, we will begin with the formal solution~(\ref{Eq:full_solution}) and evaluate the integral using the Cauchy theorem. In the short-time regime analyzed in Sec.~\ref{Sec: Bound state}, we will begin with the Dyson series~(\ref{Eq. Dyson series}). 

    We show that the short-time dynamics is still exponential and independent of the Dirac gap \( 2m \), whereas the long-time dynamics is non-exponential and depends on the Dirac gap. The pole contribution to the wave function yields a bound state, whereas the branch cut contributes to the time-evolving state. Furthermore, by closing the Dirac gap \( 2m \rightarrow 0 \), we will observe a convergence of the time-evolving state into the time-evolving resonant state that we found in Sec.~\ref{sec: L infinite and m zero}.

    \subsubsection{Survival probability: Long-time dynamics}\label{Sec: short-time}
    %Setup
    To analyze the long-time dynamics, we start with the formal solution~(\ref{Eq:full_solution}) and take the limit $\Lambda\rightarrow\infty$ with $m>0$. In this case, the arctangent function in the integrand of Eq.~(\ref{Eq:full_solution}) converges to \(\pi/2\), yielding a simplified expression for the probability amplitude
    \begin{align}
        \Phi(t)& = e^{\frac{i}{\hbar} \omega_0 t} \lim_{R \rightarrow \infty} \int_{-iR + \sigma}^{iR + \sigma} \frac{dz}{2\pi i} e^{z \frac{g^2}{\hbar} t}  F(z)\Phi(0),\label{Eq:Simpler_solution}\\
        &F(z)= \left(z + i \frac{\omega_0}{g^2} + \frac{2\pi z}{\sqrt{z^2 + m^2 / g^4}} \right)^{-1} .\label{Eq:Simpler_solution_2}
    \end{align}
    Here, we have taken the single-valued arctangent function defined on the principal branch in Eq.~(\ref{Eq:full_solution}), before taking the limit $\Lambda\rightarrow\infty$. 
    
\begin{figure*}[]
        \centering
        \includegraphics[width=0.9\textwidth]{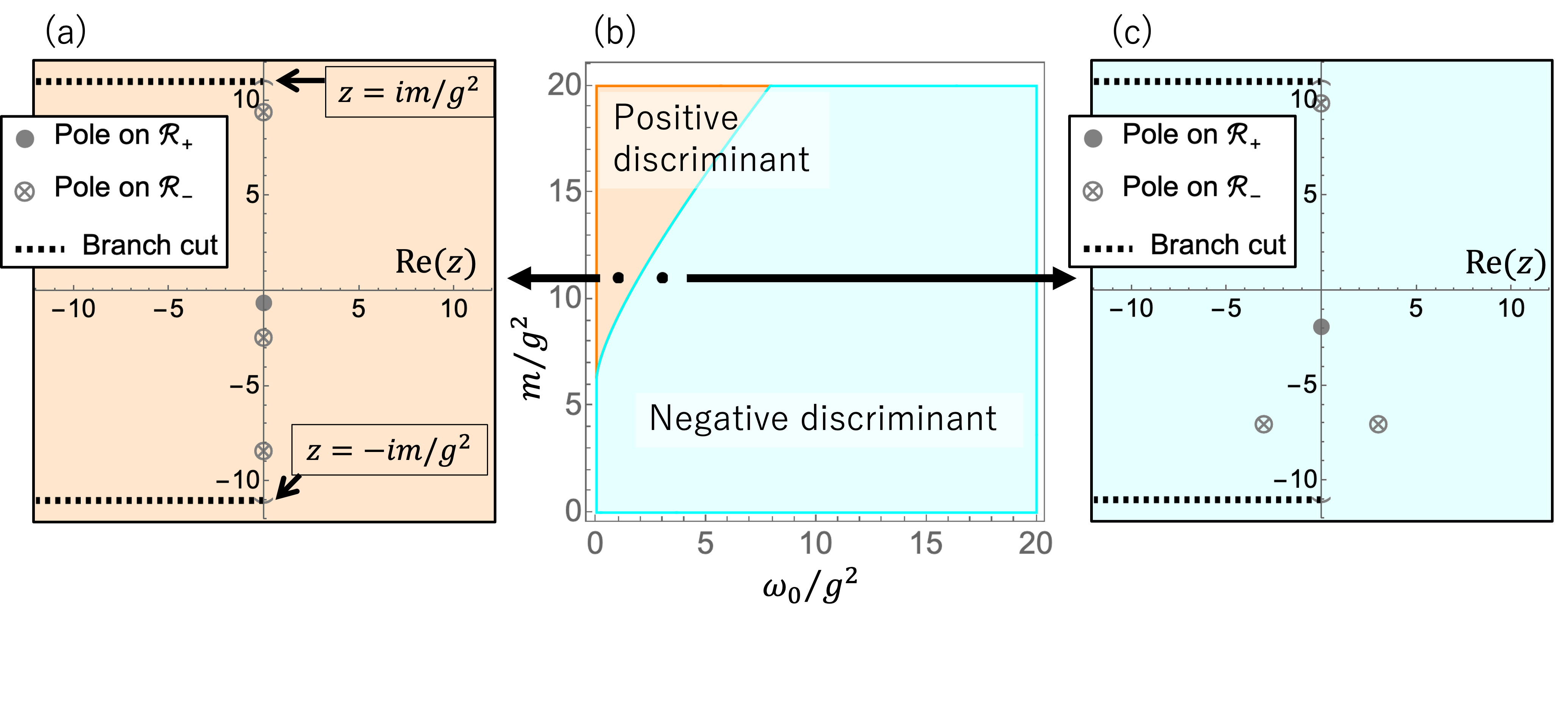}
        \caption{Panel (b) shows parameter regions where the discriminant of the polynomial equation \( \mathcal{P}(z)=0 \) is either positive or negative. Panels (a) and (c) show the corresponding locations of the poles on the first and second Riemann sheets of the function \( F(z) \), indicated by solid and crossed dots, respectively. The horizontal dashed lines in panels (a) and (c) represent the branch cuts of the function \( F(z) \).}
        \label{fig:Contour_L_infinite}
    \end{figure*}

    Note that the integrand in Eq.~(\ref{Eq:Simpler_solution}) contains a square root of the complex variable \( z \), making it a multivalued function defined on a Riemann surface. 
    In this case, the Riemann surface consists of two sheets, \( \mathcal{R}_\pm \), which correspond to the two branches of the square root. 
    These sheets are defined over the complex plane \( \mathbb{C} \) with branch cuts. 
    We choose branch cuts that extend from two branch points \( z = \pm i m / g^2 \) to negative infinity, as illustrated by the dotted lines in panels (a) and (c) of Fig.~\ref{fig:Contour_L_infinite}. 
    For convenience, we define single-valued analytic functions on each sheet:
    \begin{align}\label{Eq: integrand F}
        F_\pm (z) := \left(z + i \frac{\omega_0}{g^2} + \frac{2\pi z}{\pm \sqrt{z^2 + m^2 / g^4}} \right)^{-1}, \quad z \in \mathcal{R}_{\pm}.
    \end{align}
    
    We can evaluate the integral in Eq.~(\ref{Eq:Simpler_solution}) using Cauchy's residue theorem, reducing it to a sum of the residues of the integrand on the first Riemann sheet and the contributions from the branch cuts. 
    In the following, we will find the contributions from the poles and branch cuts of the integrand.
    
    To find the poles, we solve \( F_+(z)^{-1} = 0 \) for poles on \( \mathcal{R}_+ \) and \( F_-(z)^{-1} = 0 \) for poles on \( \mathcal{R}_- \). 
    A straightforward transformation involving squaring yields the same quartic polynomial:
    \begin{align}\label{Eq. 4th order poly}
        \mathcal{P}(z) := \left(z + i \frac{\omega_0}{g^2} \right)^2 \left( \frac{m^2}{g^4} + z^2 \right) - 4\pi^2 z^2 = 0.
    \end{align}
    Analyzing the discriminant $\Delta$ of the polynomial \( \mathcal{P}(z) \) reveals the nature of its roots. Its explicit form is written as
    \begin{align}
        &\Delta = -64 \pi^2 a \left[b_-^3 + 48\pi^4 b_- + 12\pi^2 (b_+^2 + 5a)+64\pi^6\right],\label{Eq: discriminant}\\
        &a= (\omega_0 /g^2)^2(m/g^2)^2 , \quad b_\pm = (\omega_0 /g^2)^2 \pm(m/g^2)^2.
    \end{align}
    When the discriminant $\Delta$ is negative, two of the poles are purely imaginary, while the other two are symmetric about the imaginary axis. 
    This configuration is shown in Fig.~\ref{fig:Contour_L_infinite}(c).
    We let $z_0 , z_1 \in i \mathbb{R}$ denote the purely imaginary roots and $z_\pm$ the symmetric pair, where \( z_+ = - (z_-)^* \).  
    As the discriminant $\Delta$ becomes positive, the symmetric pair \( z_\pm \) merges and splits into purely imaginary poles, as illustrated in Fig.~\ref{fig:Contour_L_infinite}(a).
    
    To determine which Riemann sheet each pole lies on, we substitute the solutions of Eq.~(\ref{Eq. 4th order poly}) into each of \( F_\pm(z)^{-1} \). 
    If a pole satisfies \( F_+(z)^{-1} = 0 \), it lies on \( \mathcal{R}_+ \); otherwise, it lies on \( \mathcal{R}_- \). 
    From this analysis, we find that only one pole—a purely imaginary and negative one denoted by \( z_0 \), namely a bound state—resides on \( \mathcal{R}_+ \). 
    The other three poles are located on \( \mathcal{R}_- \), as depicted in Fig.~\ref{fig:Contour_L_infinite}.
    
    We are now in a position to evaluate the Bromwich integral in Eq.~(\ref{Eq:Simpler_solution}). 
    The original contour, along with the locations of the poles and the branch cuts, is shown in Fig.~\ref{fig:Poles}(a). 
    According to Cauchy's theorem (see Appendix~\ref{Appdx: Contour}), we can deform this contour to the one in Fig.~\ref{fig:Poles}(b). 
    As a result, the Bromwich integral can be decomposed into two distinct contributions: one from the bound-state pole \( z_0 \) on the first Riemann sheet $\mathcal{R}_+$ and the other from the two branch cuts:
    \begin{align}\label{Eq:L_infity_exact_solution}
        \Phi(t) &=  \left( \oint_{z_0} \cdots + \int_{\text{BC}} \cdots \right) \Phi(0) \nonumber\\
        &=: \Phi_{z_0}(t) + \Phi_{\text{BC}}(t),
    \end{align}
    where \( \Phi_{z_0}(t) \) and \( \Phi_{\text{BC}}(t) \) denote the contributions from the pole and the branch cuts, respectively.

    \begin{figure}[t]
        \centering
        \includegraphics[width=0.45\textwidth]{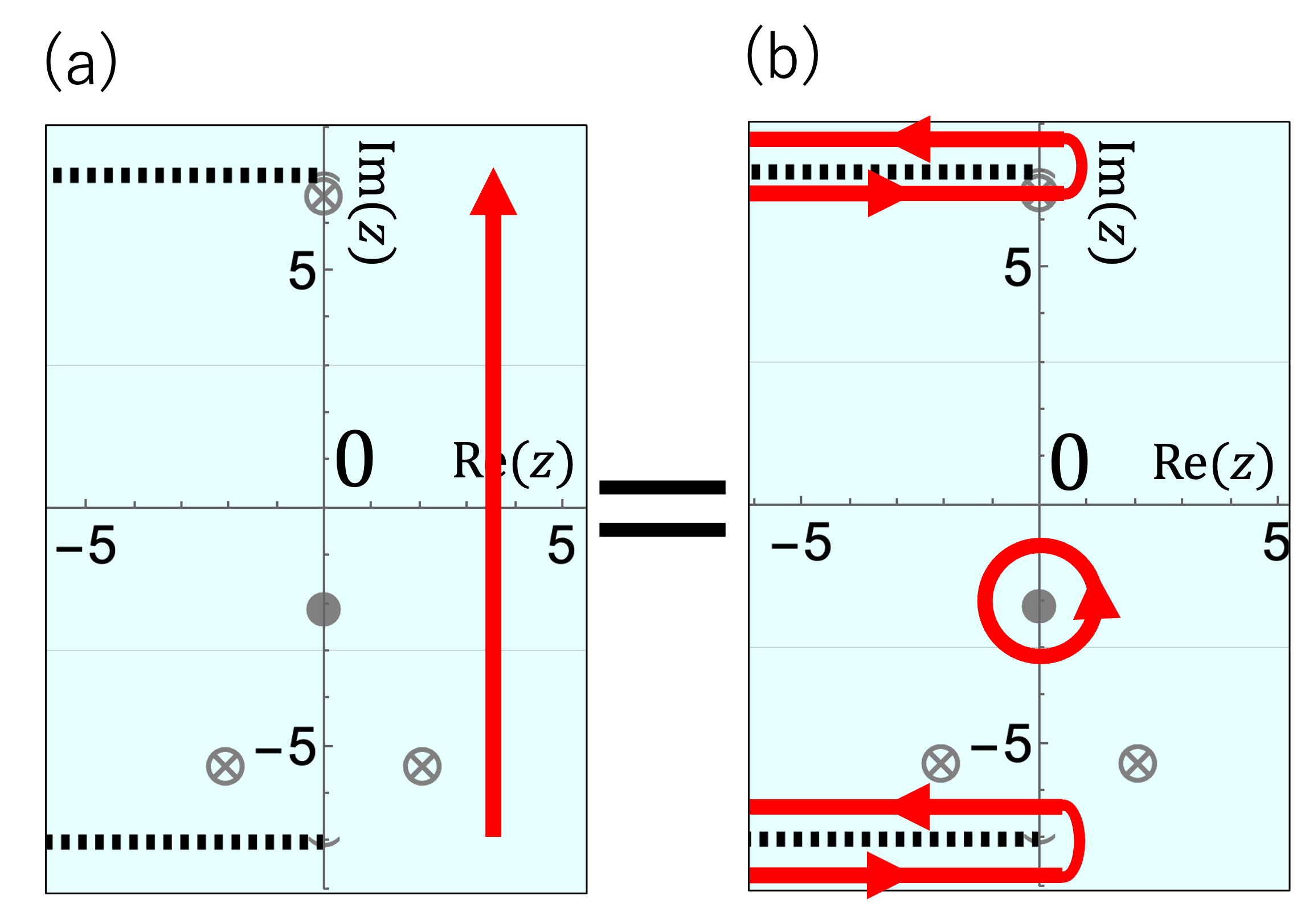}
        \caption{(a) The Bromwich contour of the integral in Eq.~(\ref{Eq:Simpler_solution}) on the complex \( z \)-plane, with parameters \( \omega_0 = 3/11 \), \( g = \sqrt{1/11} \), and \( m = 1 \) for example. (b) A deformed contour, equivalent to (a) by Cauchy's theorem, used in Eq.~(\ref{Eq:L_infity_exact_solution}).}
        \label{fig:Poles}
    \end{figure}

\begin{figure*}[t]
        \centering
        \includegraphics[width=0.7\textwidth]{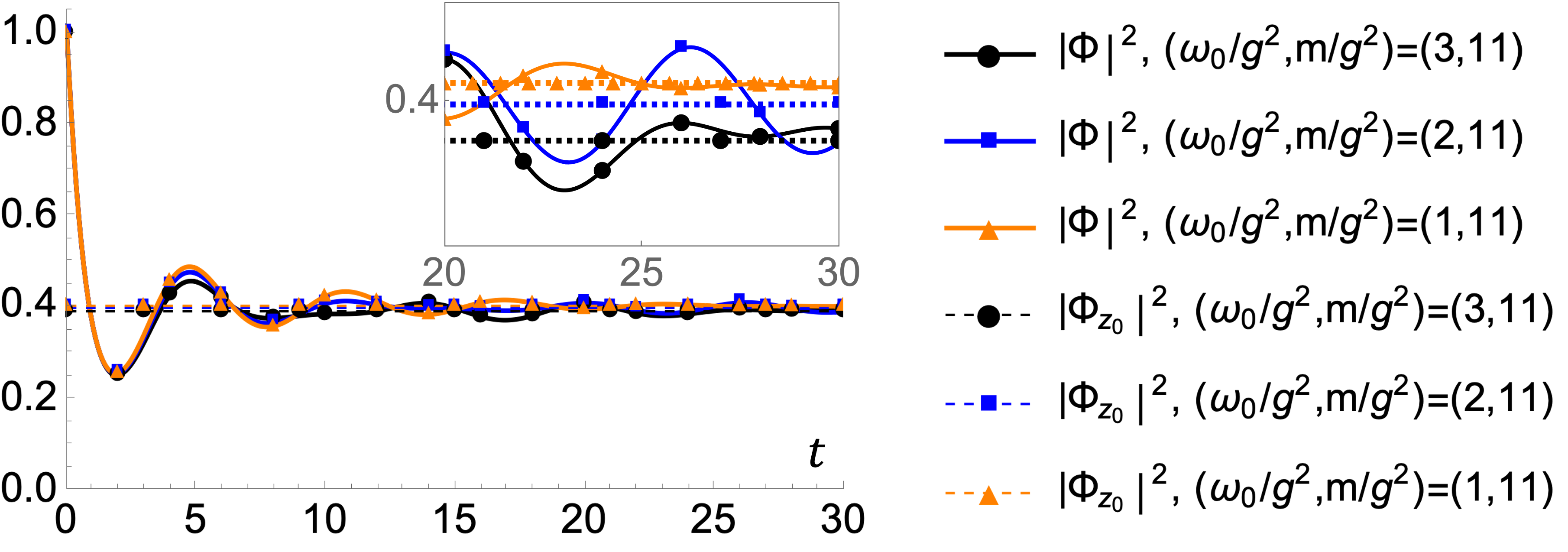}
        \caption{Survival probability in the limit \( \Lambda \rightarrow \infty \). Solid lines show the full survival probability given by Eq.~(\ref{Eq:L_infity_exact_solution}), while dashed lines indicate the contribution from the pole \( z_0 \). Different markers correspond to different parameter regimes. The inset shows the detail of a part of the main plot.}
        \label{fig: Bound state SP and EP}
    \end{figure*}

    We can evaluate the pole contribution in Eq.~(\ref{Eq:L_infity_exact_solution}), using Cauchy's residue theorem for the pole \( z_0 \):
    \begin{align}\label{Eq:L_infty_bound}
        \Phi_{z_0}(t)&=\text{Res}(F(z)e^{\frac{i}{\hbar}\omega_0 t+z\frac{g^2}{\hbar}t},z_0)\nonumber \\  
        &=  \frac{\left(z_0+i \frac{\omega_0}{g^2} \right)\left(\frac{m^2}{g^4} + z^2_0 \right)-2 \pi z_0 \sqrt{\frac{m^2}{g^4}+ z^2_0 }}{(z_0-z_1)(z_0-z_+)(z_0-z_-)} \nonumber\\
        &\times e^{\frac{i}{\hbar}\omega_0 t+z_0 \frac{g^2}{\hbar}t},
    \end{align}
    where \( z_0 \) is the pole on the Riemann sheet \( \mathcal{R}_+ \), while \( z_1 \), \( z_+ \), and \( z_- \) are the poles located on the second sheet \( \mathcal{R}_- \). 
    The absolute value squared of this expression is a constant value, plotted as the dashed horizontal lines in Fig.~\ref{fig: Bound state SP and EP} for three sets of parameter choice: \( (\omega_0 / g^2,\, m / g^2) = (1,11), (2,11), (3,11) \). 
    The solid lines shown in Fig.~\ref{fig: Bound state SP and EP} are the numerical plot of the full survival probability.

    Next, let us analyze the branch-cut contribution in Eq.~(\ref{Eq:L_infity_exact_solution}). 
    Recall the single-valued complex functions \(F_{\pm}(z)\) from Eq.~(\ref{Eq: integrand F}), which denote the integrand of Eq.~(\ref{Eq:Simpler_solution}) defined on the Riemann sheets \(\mathcal{R}_{\pm}\), respectively. 
    Let us rewrite them as
    \begin{align}
        F_\pm (z) =\frac{\left(z+i \frac{\omega_0 }{g^2} \right)\left(\frac{m^2}{g^4} + z^2 \right)}{\mathcal{P}(z)}\pm \frac{-2\pi z \sqrt{\frac{m^2 }{g^4}+ z^2 }}{\mathcal{P}(z)},
    \end{align}
    where \(\mathcal{P}(z)\) is the quartic polynomial defined in Eq.~(\ref{Eq. 4th order poly}). 
    In the square-root function in the above expression, we took the principal branch with \(\sqrt{-1} = i\).

\begin{figure}[t]
        \centering
        \includegraphics[width=0.5\textwidth]{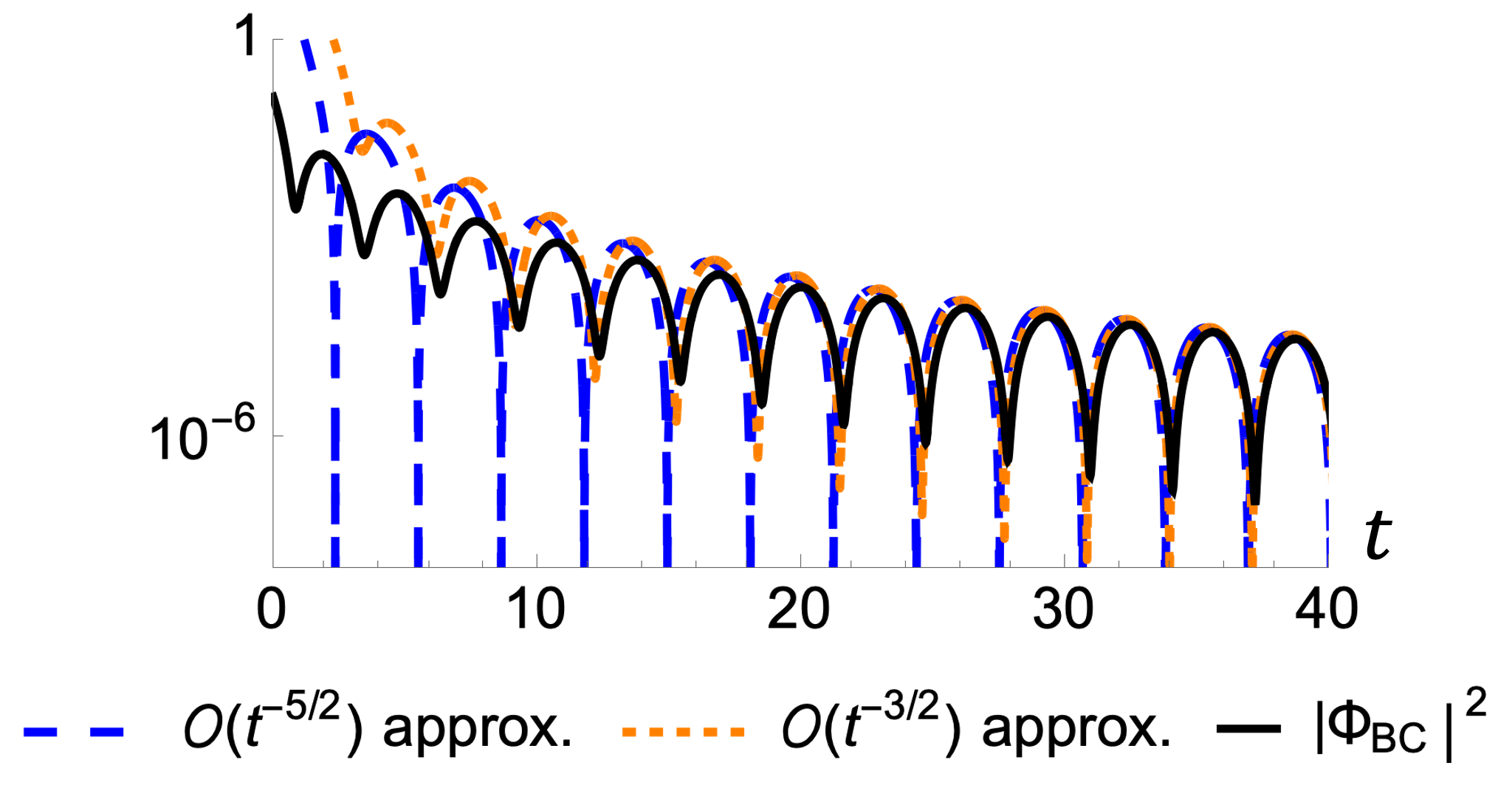}
        \caption{Logarithmic plot of the branch-cut contribution to the survival probability, \(\left|\Phi_{\text{BC}}(t)\right|^2\), compared with the asymptotic approximations up to \(\mathcal{O}(t^{-3/2})\) and \(\mathcal{O}(t^{-5/2})\) from Eq.~(\ref{Eq:L_infty_approx}). The plot is shown for parameter values \(\omega_0 = 3/11\), \(g = \sqrt{1/11}\), and \(m = 1\).}
        \label{fig:L_infty_solutions}
    \end{figure}

    Let $I_{\text{up}}$ denote the contribution of the upper branch cut in Fig.~\ref{fig:Poles}(b). It is given by
    \begin{align}
        I_{\text{up}}&=e^{\frac{i}{\hbar}\omega_0 t} \int_{-R+ i m/g^2}^{i m/g^2} \frac{dz}{2\pi i } ~  \left(F_+ (z) - F_- (z)\right)e^{z\frac{g^2}{\hbar}t}\nonumber\\
        &=  e^{\frac{i}{\hbar}\omega_0 t}  \int_{-R+ i m/g^2}^{i m/g^2}  \frac{dz}{2\pi i } ~ A(z) e^{z\frac{g^2}{\hbar}t}\nonumber\\
        &=e^{\frac{i}{\hbar}\omega_0 t} \int_{-R}^{0} \frac{dw}{2\pi i } ~ A\left(w+i\frac{m}{g^2}\right) e^{w\frac{g^2}{\hbar}t}e^{i \frac{m}{g^2}\frac{g^2}{\hbar}t},
    \end{align}
    where $A(z):= (-4  \pi z \sqrt{m^2/g^4 +z^2})/ \mathcal{P}(z)$ and we have performed the coordinate transformation $z = w + i m/g^2$. The contribution of the lower branch cut can be calculated similarly, and thus we find the full branch-cut contribution in the following form:
    \begin{align}\label{Eq:L_infinite_BC_solution}
        \Phi_{\text{BC}}(t)& =  e^{\frac{i}{\hbar}\omega_0 t}\lim_{R\rightarrow\infty}\int_{-R}^{0} \frac{dw}{2\pi i } ~ e^{w\frac{g^2}{\hbar}t}\nonumber\\
        &\times \left[A\left(w+ i m/g^2\right)e^{i \frac{m}{\hbar}t}-A\left(w- i m/g^2\right)e^{-i\frac{m}{\hbar}t}\right].
    \end{align}
    Although the analytic expression of the above integral is unknown, we can compute it numerically accurately as shown in Fig.~\ref{fig:L_infty_solutions} as a solid line.

    For the long-time dynamics of the survival probability, we can take the asymptotic limit of the branch-cut contribution~(\ref{Eq:L_infinite_BC_solution}) by following the analysis outlined in Ref.~\cite{garmon2013amplification}. 
    First, we change the complex variable from $w$ to $\tau$, as in $w t \equiv \tau$. We then expand the integrand as a series in powers of $\hbar /(g^2 t)$. 
    After evaluating the integral term by term, we find
    \begin{align}\label{Eq:L_infty_approx}
        \Phi_{\text{BC}}(t) &= -e^{\frac{i}{\hbar}\omega_0 t}\sqrt{\frac{g^2}{4m\pi^3}}\left(\frac{\hbar}{g^2 t}\right)^{3/2}\nonumber\\
        & \quad \times \left[\cos\left(\frac{m }{\hbar} t\right)+\sin\left(\frac{m }{\hbar} t\right)\right]+ \mathcal{O}\left(\left(\frac{\hbar}{g^2 t}\right)^{5/2}\right).
    \end{align}
    Therefore, we find that the asymptotic contribution from the branch cuts produces a power-law decay with oscillation frequency $m/2\pi \hbar$. 
    Furthermore, in the long-time regime, the leading-order term depends on the Dirac gap $2m$. In the next subsection~\ref{Sec: Bound state}, we show that in contrast to the long-time regime, the leading-order term of the short-time regime does not depend on the Dirac gap. Note that the limit $m\to 0$ cannot be taken term by term in this asymptotic series, since each coefficient diverges; instead, the massless limit must be taken at the level of the exact branch-cut integral before the large-$t$ expansion.
    
    The long-time dynamics are shown in Fig.~\ref{fig:L_infty_solutions}, where we have numerically plotted the full branch-cuts contribution given by Eq.~(\ref{Eq:L_infinite_BC_solution}) along with the asymptotic approximations up to order $\mathcal{O}(t^{-3/2})$ and $\mathcal{O}(t^{-5/2})$, as derived in Eq.~(\ref{Eq:L_infty_approx}). 
    We observe good agreement with the full solution beyond $t = 10$, demonstrating that the asymptotic analysis provides an accurate prediction in this time regime.  

    The Markovianity in terms of the semigroup property is also discussed in Appendix~\ref{Appdx: CP-divisibility}, where we observed that the semigroup property is explicitly broken by using the explicit expression of the probability amplitude~(\ref{Eq:L_infty_approx}), indicating non-Markovian dynamics in this regime (see Appendix~\ref{Appdx: CP-divisibility}).

    \subsubsection{Survival probability: Short-time dynamics}\label{Sec: Bound state}
    
    For the analysis of the short-time dynamics of the survival probability, the Dyson series expansion~(\ref{Eq. Dyson series}) is more convenient than the Bromwich integral~(\ref{Eq:full_solution}) used in the previous subsection~\ref{Sec: short-time}.
    
    We begin by noting that, in the limit \( \Lambda \rightarrow \infty \), the memory kernel \( \mathcal{K}(t) \) in Eq.~(\ref{Eq:Memory_kernel}) can be evaluated explicitly via the Laplace transform:
    \begin{align}\label{Eq: memoery kernel in massive case}
        \mathcal{K}(t) = - \frac{2\pi g^2}{\hbar^2} e^{\frac{i}{\hbar} \omega_0 t} \left[\hbar \delta(t ) - m J_1 \left( \frac{m t}{\hbar} \right) \right],
    \end{align}
    where \( J_1(t) \) is the Bessel function of the first kind. Note that taking the limit \( m \rightarrow 0 \) recovers Eq.~(\ref{Eq: memory kernel in Marko limit}) because $J_1(0)=0$.
    Using the expression~(\ref{Eq: memoery kernel in massive case}), we can evaluate the first few terms in the exponent of the Dyson series as
    \begin{align}
        \int_0^{t} ds \int_0^{t-s} du\, \mathcal{K}(u) &= -2\pi \left(\frac{g^2}{\hbar} t \right) + \pi \frac{(m/g^2)^2}{6} \left(\frac{g^2}{\hbar} t \right)^3 \nonumber\\
        &+ \mathcal{O}\left(\left( \frac{g^2}{\hbar} t \right)^4 \right).
    \end{align}
    We thereby obtain the short-time expansion of the solution to the non-Markovian dynamical equation~(\ref{Eq:Master equation simple form}) as follows:
    \begin{align}\label{Eq: perturbative expansion}
        \Phi(t) = \exp\left[ -2\pi \frac{g^2}{\hbar} t + \pi \frac{(m / g^2)^2}{6} \left(\frac{g^2}{\hbar} t \right)^3 + \dots \right].
    \end{align}
    Using this expansion, we arrive at the short-time behavior of the survival probability \( P(t) = |\Phi(t)|^2 \) in the form
    \begin{align}\label{Eq: short time solution}
        P(t) &= 1 - 4\pi \left(\frac{g^2}{\hbar} t \right) + 8 \pi^2 \left( \frac{g^2}{\hbar} t \right)^2 \nonumber\\
        &+ \frac{1}{3} \left[ \pi \left( \frac{m}{g^2} \right)^2 - 32 \pi^3 \right] \left( \frac{g^2}{\hbar} t \right)^3 + \mathcal{O}\left( \left( \frac{g^2}{\hbar} t \right)^4 \right).
    \end{align}
    We find that the survival probability decays linearly in the short-time regime, implying exponential decay. Furthermore, up to the second order in time, the survival probability does not depend on the Dirac gap \( 2m \), meaning that the short-time exponential behavior holds for both massive and massless cases. From the leading-order probability amplitude in Eq.~(\ref{Eq: perturbative expansion}), we also find that the dynamics is consistent with the semigroup property at the leading order in the short-time regime, indicating Markovian behavior in this approximation (see Appendix~\ref{Appdx: CP-divisibility}).
    
    This result implies that the quantum Zeno effect (see Appendix~\ref{Appdx: Quantum Zeno Effect}), which requires quadratic short-time decay, is absent for any value of the Dirac gap \( 2m \) when the momentum cutoff is absent (\textit{i.e.}, \( \Lambda \rightarrow \infty \)). We will see in the next section that quadratic decay emerges when the cutoff \( \Lambda \) is finite, suggesting that the quantum Zeno effect originates from the spectral bound rather than from the gap structure.

      Finally, summarizing our analysis of the survival probability in the short- and long-time regimes in the case of $m>0$ and $\Lambda\rightarrow\infty$, we obtain the following asymptotic expressions for the probability amplitude:
    \begin{subequations}\label{Eq: Asym-exact 1}
        \begin{align}
        \Phi (t) &\sim  \exp \left(- 2\pi\frac{g^2}{\hbar} t  \right) ~\text{ as }~\frac{g^2}{\hbar} t \rightarrow 0 ,\label{Eq: Asym-exact short 1}\\
        \Phi (t)  & \sim \Phi_{z_0} (t) -e^{\frac{i}{\hbar}\omega_0 t}\sqrt{\frac{g^2}{4m\pi^3}}\left(\frac{\hbar}{g^2 t}\right)^{3/2} \nonumber\\
        &\times \left[\cos\left(\frac{m }{\hbar} t\right)+\sin\left(\frac{m }{\hbar} t\right)\right]  ~\text{ as }~\frac{g^2}{\hbar} t \rightarrow \infty ,\label{Eq: Asym-exact long 1}
    \end{align}
    \end{subequations}
    where $\Phi_{z_0}(t)$ is the pole contribution given in Eq.~(\ref{Eq:L_infty_bound}).

    This behavior of exponential decay in the short-time regime, followed by a non-exponential decay, was previously observed in the Friedrichs-Lee model, where the coupling between the system and the environment was engineered to produce exponential decay in the short-time regime, followed by non-exponential decay. This was dubbed \textit{hidden non-Markovianity} in Ref.~\cite{burgarth2021hidden}.
    
    \subsubsection{Wave function of environmental states}\label{sec: wave function of massive no-cutoff case}
    The wave function can also be evaluated by taking the limit \( \Lambda \rightarrow \infty \) in the formal solution~(\ref{Eq: Wave function}) and explicitly calculating the second integral:
    \begin{align}
        \psi(t,x) &= \frac{1}{i} \sqrt{\frac{2g^2\pi}{\hbar}} \lim_{R \rightarrow \infty} \int_{-iR + \sigma}^{iR + \sigma} \frac{dz}{2\pi i} \nonumber\\
        &\times 
        \frac{z\, e^{\frac{g^2}{\hbar} z t} e^{- \frac{g^2}{\hbar} |x| \sqrt{z^2 + (m / g^2)^2}}}{\left(z + i \omega_0 / g^2 \right) \sqrt{z^2 + (m / g^2)^2} + 2\pi z}.
    \end{align}
    We have expressed the wave function in terms of the dimensionless parameters \( \omega_0 / g^2 \) and \( m / g^2 \). 
    Notice that the pole and the branch-cut structure is the same as for the survival probability in Eq.~(\ref{Eq:Simpler_solution}). 
    Therefore, we can use the same argument as above to decompose the Bromwich integral into the pole and the branch-cut contributions:
    \begin{align}\label{Eq: massless wave function}
        \psi(t,x) &= \frac{1}{i} \sqrt{\frac{2g^2\pi}{\hbar}} \left( \oint_{z_0} \cdots + \int_{\text{BC}} \cdots \right) \nonumber\\
        &=: \psi_{z_0}(t,x) + \psi_{\text{BC}}(t,x).
    \end{align}
    
\begin{figure}[t]
        \centering
        \includegraphics[width=0.5\textwidth]{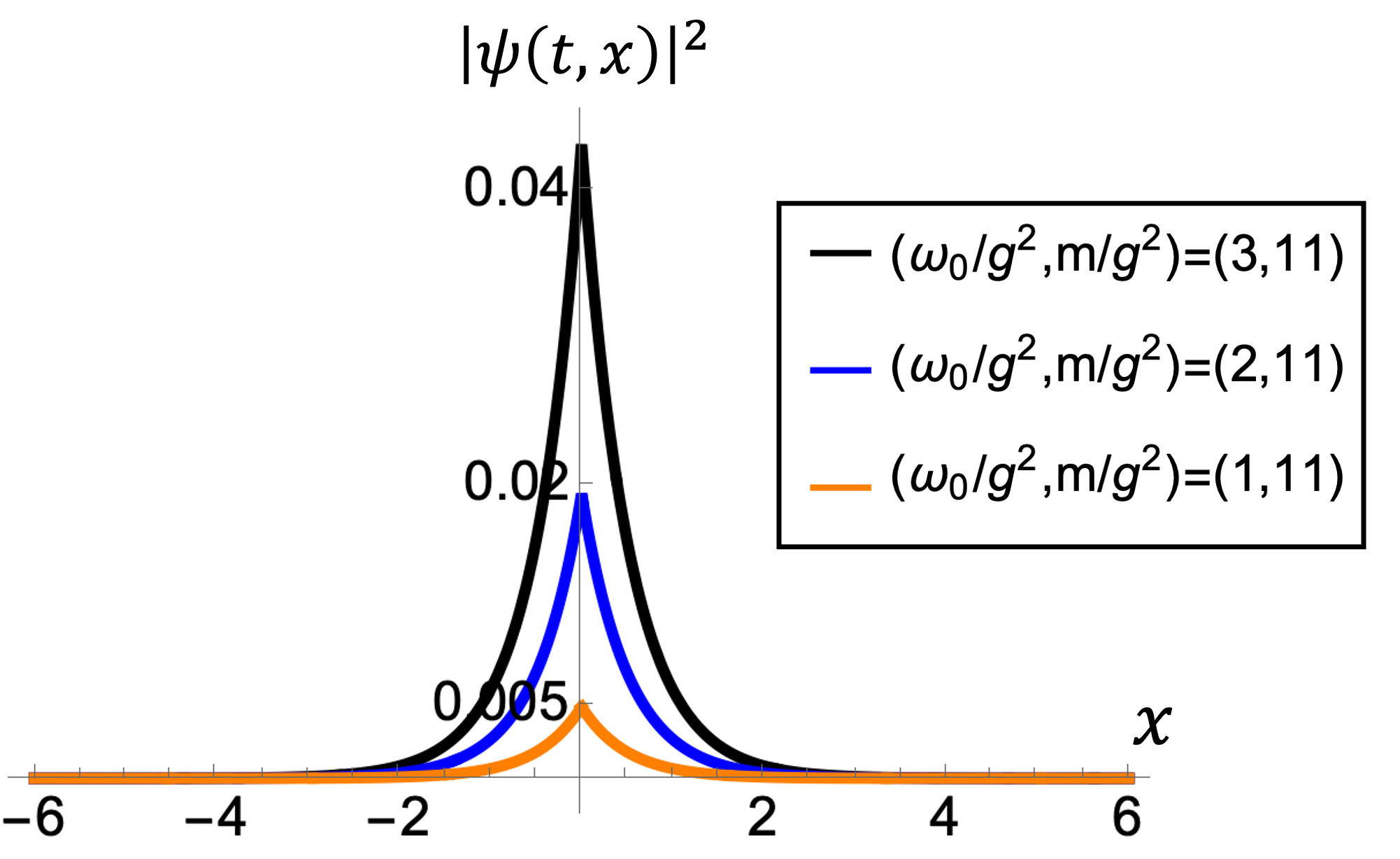}
        \caption{The probability profile of the bound state wave function, given by Eq.~(\ref{Eq: exact bound state}), plotted for three different sets of parameters. Time is taken to be $t=0$, but it is time-independent.}
        \label{fig: Bound State}
    \end{figure}

    Next, let us consider the wave function given in Eq.~(\ref{Eq: massless wave function}). We will first show that the pole \( z_0 \) indeed gives the wave function of a bound state. The pole contribution in Eq.~(\ref{Eq: massless wave function}) is given by
    \begin{align}\label{Eq: exact bound state}
        \psi_{z_0} (t,x) &= \frac{1}{i}\sqrt{\frac{2g^2\pi}{\hbar}} e^{z_0 \frac{g^2}{\hbar}t - \frac{g^2}{\hbar} \left|x\right| \sqrt{z_0^2 + m^2 / g^4}}\nonumber\\
        & \times \frac{z_0 \left(z_0 +i \frac{\omega_0}{g^2} \right)\sqrt{z_0^2 + \frac{m^2}{g^4}} - 2 \pi  z_0^2 }{(z_0 - z_1 ) (z_0 - z_+ )(z_0 - z_-)}  .
    \end{align}
    The absolute value squared \( |\psi_{z_0}(t,x)|^2 \) of this bound state is plotted in Fig.~\ref{fig: Bound State} for the same parameter sets used previously: \( (\omega_0 / g^2,\, m / g^2) = (1,11), (2,11), (3,11) \). Notice that the $x$-dependent part of the wave function is of the form $e^{-K \left|x\right|}$ where $K = g^2 \sqrt{z_0^2 + m^2 / g^4}$ is always a real positive value due to $z_0$ being a purely imaginary pole that lies between $m/g^2 > \left|z_0\right| > 0$. This gives the characteristic exponential decay in real space, consistent with the known behavior of the bound state.
    
\begin{figure*}[t]
        \centering
        \includegraphics[width=\textwidth]{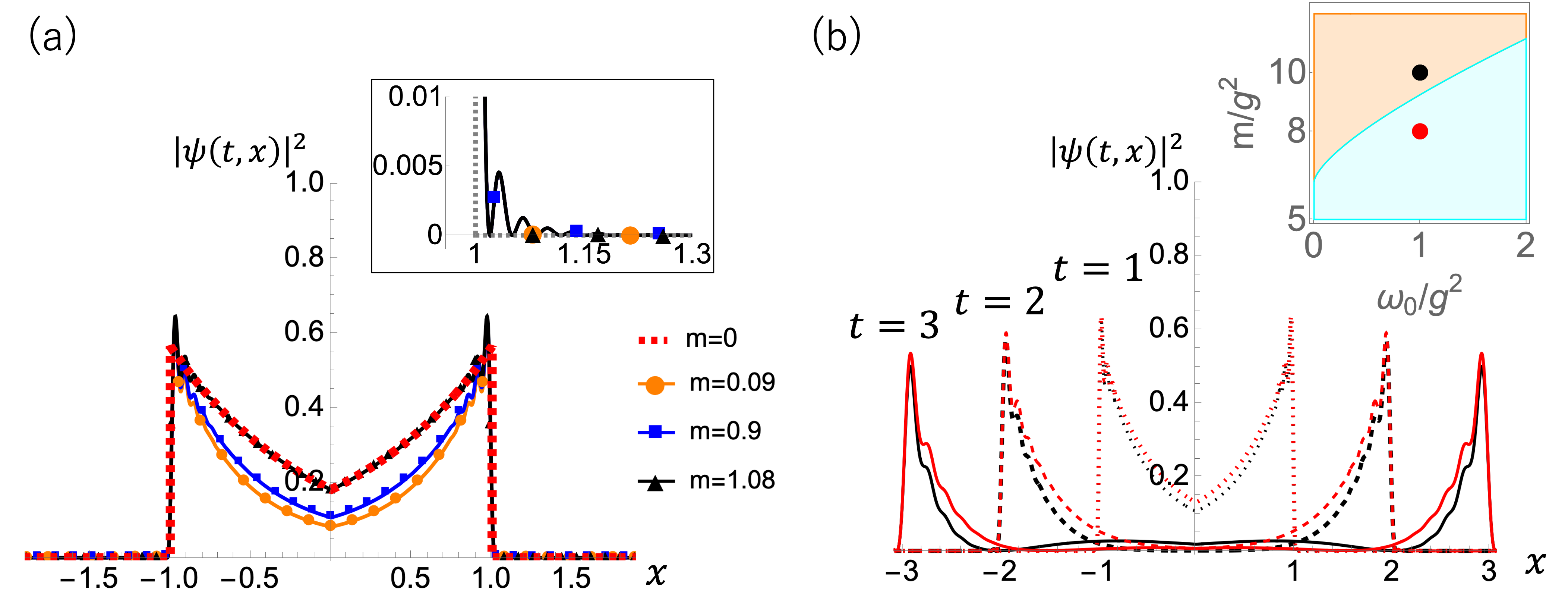}
        \caption{(a) The probability profile of the wave function~(\ref{Eq: massless wave function}), plotted for $m = 0$, $m = 0.09$, $m = 0.9$, and $m = 1.08$, with parameter values $\omega_0 = 0.09$, $g = 0.3$, and $t = 1$. The inset shows a zoomed-in region where the probability drops to zero beyond $t = 1$. (b) The probability profile of the wave function~(\ref{Eq: massless wave function}) for two different parameter sets. The black lines correspond to $g = 0.3$, $\omega_0 / g^2 = 1$, and $m / g^2 = 10$, plotted at times $t = 1$, $2$, and~$3$. The red curves correspond to $g = 0.3$, $\omega_0 / g^2 = 1$, and $m / g^2 = 8$, also at $t = 1$, $2$, and $3$. The inset shows the part of the phase diagram in Fig.~\ref{fig:Contour_L_infinite}, where black and red dots indicate the parameter values used for the black and red lines, respectively.}
        \label{fig:Massive Wave}
    \end{figure*}

    The probability density of the wave function~(\ref{Eq: massless wave function}) is plotted in Fig.~\ref{fig:Massive Wave}(a) for a coupling strength of $g = 0.3$, with the system frequency set to $\omega_0 = 0.09$ and the time fixed at $t = 1$. 
    The exponentially decaying profile of the bound state shown in Fig.~\ref{fig: Bound State} is not discernible in Fig.~\ref{fig:Massive Wave} because its amplitude is small on the scale of the full time-evolving wave function~(\ref{Eq: massless wave function}). This means that the plot in Fig.~\ref{fig:Massive Wave} visualizes mostly the branch-cut contribution $\psi_{\text{BC}}(t,x)$ of Eq.~(\ref{Eq: massless wave function}).
    
    Three solid lines are shown in panel~(a), corresponding to the Dirac mass $m = 0.09$, $0.9$, and $1.08$, while the dashed line represents the massless case $m = 0$, which corresponds to the time-evolving resonant state described in Eq.~(\ref{Eq: Exp wave function}) and in Fig.~\ref{fig:Exp_decay_and_wave}. Note that, three wave functions for $m = 0.09$, $0.9$, and $1.08$ can be interpreted as an approximate time-evolving resonant state as it approaches the exact time-evolving resonant state in the limit $m\rightarrow 0$. 
    
    The oscillations near the causal point $t = |x|$ become increasingly vibrant as the Dirac mass approaches zero. 
    This behavior arises because, for a small value of the Dirac mass, the Laplace transform in the wave function~(\ref{Eq: massless wave function}) approximates the step function that appears in the massless case in Eq.~(\ref{Eq: Exp wave function}). 
    Such oscillatory behavior is also characteristic of the standard Fourier approximation of a step function, known as the Gibbs phenomenon.
    
    In Fig.~\ref{fig:Massive Wave}(b), we plot the probability density of the wave function~(\ref{Eq: massless wave function}) at three different times $t=1,2,3$. 
    Two sets of parameter values are chosen such that the discriminant~(\ref{Eq: discriminant}) of the corresponding quartic polynomial may be either positive or negative. 
    Specifically, the parameters $g = 0.3$, $\omega_0 / g^2 = 1$, and $m / g^2 = 10$ correspond to a positive discriminant, while $g = 0.3$, $\omega_0 / g^2 = 1$, and $m / g^2 = 8$ yield a negative discriminant. 
    This distinction determines the locations of the poles, as discussed in Sec.~\ref{Sec: short-time} and illustrated in Fig.~\ref{fig:Contour_L_infinite}. 
    For a positive discriminant, all four poles lie on the imaginary axis, with only one located on the first Riemann sheet. 
    For a negative discriminant, two of the three poles on the second Riemann sheet acquire nonzero real parts.
    These complex poles are commonly referred to as the resonant and anti-resonant poles~\cite{Hatano2026resonant}.
    (We note that this is not the same as the time-evolving resonant state observed in~Sec.~\ref{sec: L infinite and m zero}.)
    
    Let us comment on the exceptional point that arises in our model. 
    The boundary between the positive and negative discriminant shown in Fig.~\ref{fig:Contour_L_infinite}(b) is often referred to as the exceptional point of the open quantum system~\cite{rotter2009non}, which is the non-diagonalizable parameter point of the Feshbach effective Hamiltonian of the open system because the two poles coalesce. 
    A characteristic change in the dynamics of the system is often observed at the exceptional point. 
    However, from Fig.~\ref{fig:Massive Wave}(b), we observe no characteristic change by crossing the exceptional point. 
    Previous studies employing the tight-binding model have shown that the physical effects of crossing this point is not immediately evident~\cite{garmon2017characteristic}. 
    This suggests that the effect of the exceptional point in an open quantum system is a subtle effect and requires further investigation that is left for future work.

\subsection{High energy structure: $m=0$ and $ \Lambda<\infty$}\label{Sec:High_energy_structure_and_Markovianity}
    The main objective of this subsection is to examine how the finite spectral cutoff $\Lambda < \infty$ affects the exponential decay and the time-evolving resonant state identified in Sec.~\ref{sec: L infinite and m zero}. To isolate this effect, we set $m = 0$, thereby removing low-energy contributions. Similar to the previous section~\ref{Sec:Low_energy_structure_and_Markovianity}, we use different approaches to investigate the long-time regime in Sec.~\ref{sec: short time with finite cut-off} and the short-time regime in Sec.~\ref{Sec: short time with finite cut-off}. We investigate the wave function in Sec.~\ref{sec: long time with finite cut-off}.
    
    We show that both short- and long-time dynamics are non-exponential and depend on the cutoff $\Lambda < \infty$. The branch-cut contribution to the wave function yields a time-evolving state. Furthermore, by taking the limit $\Lambda \rightarrow \infty $, we will once again observe convergence of the time-evolving state into the time-evolving resonant state, similar to the one that we observed in Sec.~\ref{sec: wave function of massive no-cutoff case}.

    \subsubsection{Survival probability: Long-time dynamics}\label{sec: short time with finite cut-off}
        To analyze the long-time dynamics, we start with the formal solution~(\ref{Eq:full_solution}) and take the limit $m\rightarrow0$ while keeping a finite cutoff $\Lambda$. In this case, the formal solution given in Eq.~(\ref{Eq:full_solution}) simplifies to
\begin{align}
    \Phi(t) &= e^{\frac{i}{\hbar}\omega_0 t} \lim_{R\rightarrow \infty}\int_{-i R + \sigma}^{i R+ \sigma} \frac{dz}{2\pi i}e^{\frac{g^2}{\hbar}  z t} F(z)\Phi(0)\label{Eq:L_finite_formal_solution},\\
    &F(z)= \left[z+i \frac{\omega_0}{g^2}+ 4
    \text{arctan}\left( \frac{\Lambda / g^2}{z }\right)\right]^{-1} \label{Eq:L_finite_formal_solution_2} ,
\end{align}
where $\text{arctan}(z)$ is a multi-valued complex arctangent function. 
Recall that we have rescaled the integration variable to express the integrand in terms of the dimensionless parameters $\omega_0 / g^2$ and $\Lambda / g^2$.

\begin{figure}[ht]
	\centering
	\includegraphics[width=0.4\textwidth]{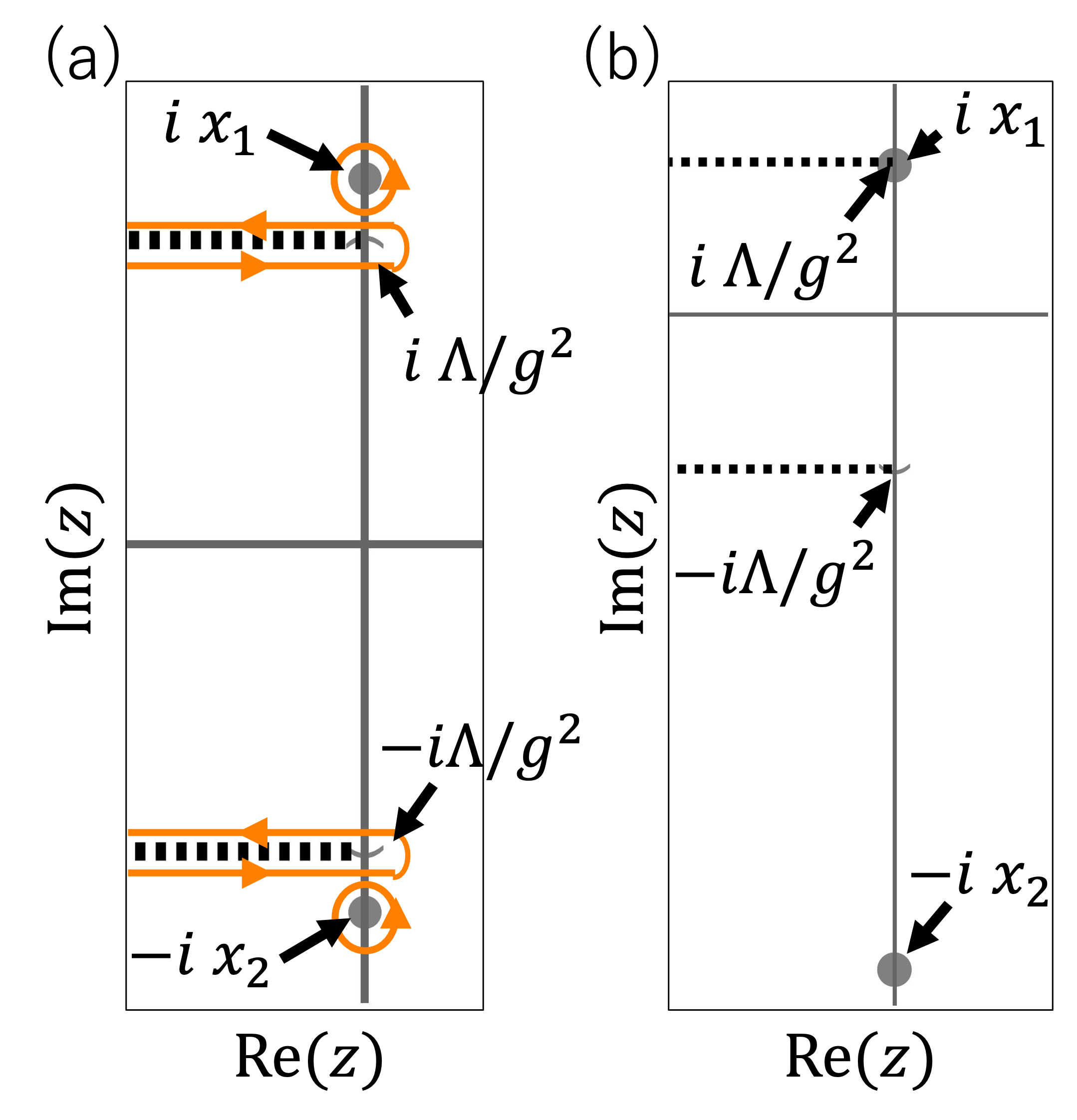}
	\caption{Panels (a) and (b) show the plots of poles on the principal Riemann surface for \(\Lambda / g^2 = 5\), \(\omega_0 / g^2 = 1\), and \(\Lambda / g^2 = 5\), \(\omega_0 / g^2 = 10\), respectively. The infinite number of poles on the other Riemann sheets is not shown in this figure.}
	\label{fig: L_finite_poles}
\end{figure}

The integrand now has a different structure of Riemann surface compared to the low-energy case discussed in Sec.~\ref{Sec:Low_energy_structure_and_Markovianity}. The branch points are now located at $z = \pm i \Lambda / g^2$ and at infinity. To find the poles, we solve the following equation:
\begin{align}\label{Eq:L_finite_pole_equation}
    0&=z + i \frac{\omega_0}{g^2} + 4\text{arctan}\left( \frac{\Lambda / g^2}{z} \right) \nonumber\\
    &= z + i \frac{\omega_0}{g^2} - i 2\text{Log}\left( \frac{z+i \Lambda/g^2}{z-i \Lambda/g^2} \right).
\end{align}
This transcendental equation cannot yield solutions of closed forms. However, by defining $w = -i (z/2) - (\Lambda / 2g^2)$, $r = -\exp[-(\omega_0 / 2g^2)- (\Lambda / 2g^2)]$, and $\alpha = 2  r (\Lambda / 2g^2)$, we can rewrite it as
\begin{align}
    w e^w + r w = \alpha, \quad r \in \left[-1,0\right), \quad \alpha \in \left[- \Lambda / g^2, 0\right).
\end{align}
This is a reformulated version of the generalized Euler–Lambert or $r$-Lambert equation~\cite{mezHo2017structure}. 
The solution $w$ to this equation is given by the $r$-Lambert function of the parameter $\alpha$, written as $W_r(\alpha)$. 
The branch structure of this function has been rigorously analyzed in Ref.~\cite{mezHo2017structure}, which shows that there are only two real solutions on the principal Riemann sheet, corresponding to purely imaginary solutions in the $z$ plane, as shown in Fig.~\ref{fig: L_finite_poles}. 
These two real solutions are the only ones found on the principal Riemann sheet for $\alpha \in \left[- \Lambda / g^2, 0\right)$.

From the above analysis of the Riemann surface, we find two purely imaginary poles, $i x_1$ and $-i x_2$, where $x_1, x_2 > \Lambda / g^2$, and the two branch cuts of the integrand in Eq.~(\ref{Eq:L_finite_formal_solution}). 
Two examples of the pole and branch point locations are shown in Fig.~\ref{fig: L_finite_poles}(a) for parameter values $\Lambda = 2$, $g = 1$, and $\omega_0 = 0$, and in Fig.~\ref{fig: L_finite_poles}(b) for $\Lambda = 2$, $g = 1$, and $\omega_0 = 8$.
Notice that the poles are almost in complex conjugate pairs in Fig.~\ref{fig: L_finite_poles}(a), whereas they are located asymmetrically in Fig.~\ref{fig: L_finite_poles}(b).

Following the same approach as the one in the previous subsection~\ref{Sec: short-time}, we can rewrite the Bromwich integral as a sum of contributions from the poles and the branch cuts:
\begin{align}\label{Eq: L finite exact solution}
    \Phi(t) &= \left[\oint_{ix_1}\dots+\oint_{-ix_2}\dots+\int_{\text{BC}}\dots\right]\Phi(0) \nonumber\\
    &=:  \Phi_{ix_1}(t) + \Phi_{-ix_2} (t)+ \Phi_{\text{BC}}(t).
\end{align}

The contribution of the poles in Eq.~(\ref{Eq: L finite exact solution}) is found by the residue theorem. The upper-pole contribution is given by
        \begin{align}
            \Phi_{ix_1} (t)&= \text{Res}\left[F(z)e^{\frac{i}{\hbar} \omega_0 t + z\frac{g^2}{\hbar} t}, i x_1\right] \nonumber\\
            &= \lim_{z \to i x_1} \frac{(z - i x_1) e^{\frac{i}{\hbar}\omega_0 t + zt}}{z + i \omega_0/g^2 + 4\text{Arctan}\left( \Lambda / z g^2 \right)},
        \end{align}
        where $F(z)$ is defined by Eq.~(\ref{Eq:L_finite_formal_solution_2}).
        Evaluating this limit is challenging because the denominator cannot be expressed in a polynomial form. To resolve this, we apply l'Hôpital's rule, which equates the limit of a fraction to the limit of the fraction of its derivatives, thereby simplifying the calculation of the residue:
        \begin{align}
            \Phi_{ix_1} (t) &= \lim_{z \to i x_1}\frac{(z - i x_1) e^{\frac{i}{\hbar}\omega_0 t +zt}}{z + i \omega_0 / g^2 + 4\text{Arctan}\left( \Lambda / z g^2 \right)}\nonumber\\
            &= \frac{x_1^2 - (\Lambda/g^2)^2}{x_1^2 + \left(4 - \Lambda/g^2\right) \Lambda/g^2}e^{\frac{i}{\hbar}\omega_0 t +i \frac{g^2}{\hbar}  x_1 t}.
        \end{align}
        
        Similarly, the contribution from the lower pole is determined using l'Hôpital's rule again. 
        We derive the combined contribution of both poles, $\text{Pole}(t) =: \Phi_{ix_1} (t)+ \Phi_{-ix_2}(t)$, as
        \begin{align}\label{Eq: Pole contribution}
            \text{Pole}(t) &=\frac{x_1^2 - (\Lambda/g^2)^2}{x_1^2 + \left(4 - \Lambda/g^2\right) \Lambda/g^2}e^{\frac{i}{\hbar}\omega_0 t +i \frac{g^2}{\hbar}  x_1 t} 
            \nonumber\\
            &+ \frac{x_2^2 - (\Lambda/g^2)^2}{x_2^2 + \left(4 - \Lambda/g^2\right) \Lambda/g^2}e^{\frac{i}{\hbar}\omega_0 t -i \frac{g^2}{\hbar}  x_2 t}.
        \end{align}

        Notice that when the two poles $i x_1$ and $-i x_2$ are almost complex conjugates of each other, as shown in Fig.~\ref{fig: L_finite_poles}(a) (\textit{i.e.}, $x_1 \sim x_2$), the above double-pole contribution approximately produces a periodic form:
        \begin{align}\label{Eq: L finite periodic}
            \text{Pole}(t) \sim 2e^{\frac{i}{\hbar}\omega_0 t}\frac{x_1^2 - (\Lambda/g^2)^2 }{x_1^2 + \left(4 - \Lambda/g^2\right)\Lambda/g^2}\cos\left(\frac{g^2 x_1}{\hbar}t\right).
        \end{align}
        This corresponds to the Rabi oscillation between the two bound states. On the other hand, when one of the pole is close to the branch point and the other is far from it, as shown in Fig.~\ref{fig: L_finite_poles}(b), the pole contribution is dominated by a single pole, as in
        \begin{align}\label{Eq: L finite bound}
            \text{Pole}(t) \sim e^{\frac{i}{\hbar}\omega_0 t} \frac{x_2^2 - (\Lambda/g^2)^2}{x_2^2 + \left(4 - \Lambda/g^2\right) \Lambda/g^2}e^{-i \frac{g^2}{\hbar}  x_2 t},
        \end{align}
        where we assume $\left|x_2\right|\gg \left|x_1\right|$.
        
        \begin{figure*}[t]
        	\centering
        	\includegraphics[width=1\textwidth]{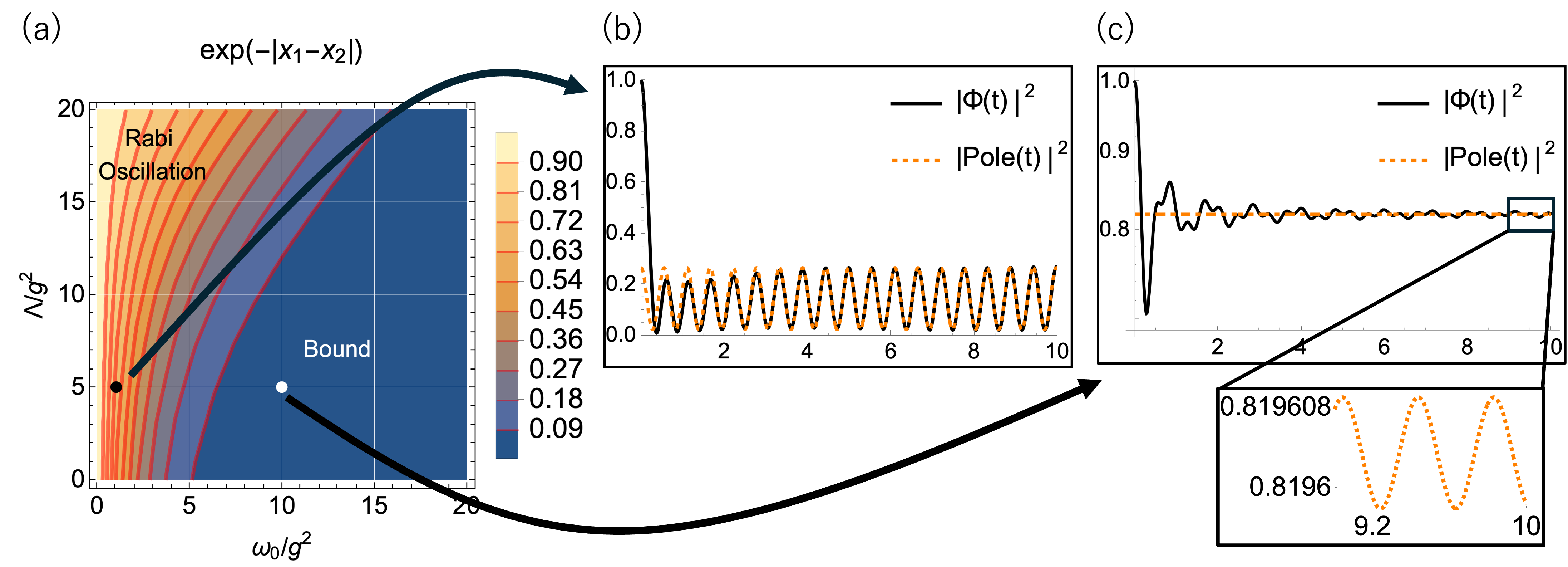}
        	\caption{(a) Contour plot of $\exp(-|x_1 - x_2|)$, showing the Rabi oscillation region, where $x_1 \sim x_2$, and the bound state region, where $x_1$ and $x_2$ are far apart. Two points indicate the parameter choices for the survival probability plots shown in panels (b) and (c). (b,c) Survival probability plots for the parameter choices indicated in panel (a). The solid line represents the numerical plot of the amplitude from Eq.~(\ref{Eq:L_finite_formal_solution}), and the dashed line represents the pole contribution from Eq.~(\ref{Eq: L finite periodic}).}
        	\label{fig: L_finite_Contour}
        \end{figure*}
        
        The above analysis reveals that, depending on the distances from each pole of $i x_1$ and $-i x_2$ to the branch points $i \Lambda / g^2$ and $-i \Lambda / g^2$, one observes a smooth crossover from the Rabi oscillation between the two bound states to the decay to the single bound state. To investigate the parameter regimes, we plot the contour of the function $e^{-|x_1 - x_2|}$ over the parameters $\omega_0 / g^2$ and $\Lambda / g^2$, as shown in Fig.~\ref{fig: L_finite_Contour}(a). 
        Notice the smooth crossover from the bright region to the dark region.
        
        The bright region corresponds to the values of $e^{-|x_1 - x_2|}$ close to unity, indicating that the two poles $i x_1$ and $-i x_2$ are close to each other, which corresponds to the Rabi oscillation. 
        As an example, we select a point $(\omega_0 / g^2, \Lambda / g^2) = (1, 5)$ in the bright region. 
        The corresponding survival probability is shown in Fig.~\ref{fig: L_finite_Contour}(b), where it exhibits an oscillation in the long-time regime.
        
        The dark region corresponds to the values of $e^{-|x_1 - x_2|}$ close to zero, indicating that the two poles $i x_1$ and $-i x_2$ are far apart from each other, which corresponds to decay to the single bound state. 
        As an example, we select a point $(\omega_0 / g^2, \Lambda / g^2) = (10, 5)$ in the dark region. 
        The corresponding survival probability is shown in Fig.~\ref{fig: L_finite_Contour}(c), where it appears to show decay to the bound state in the long-time regime. 
        However, upon zooming into the survival probability in the long-time regime, a small oscillation between the bound states is still visible, as shown in the inset of Fig.~\ref{fig: L_finite_Contour}(c). 
        The amplitude of this oscillation decreases as we select parameter points further from the bright region.
        
        Next, we consider the branch-cut contribution. 
        As we will show through the asymptotic analysis, the non-exponential decay of the survival probability in the long-time regime originates from the branch-cut term.  
        To analyze this contribution, we follow the same procedure outlined in Sec.~\ref{Sec:Low_energy_structure_and_Markovianity}.
        
        \begin{figure*}[t]
        	\centering
        	\includegraphics[width=0.7\textwidth]{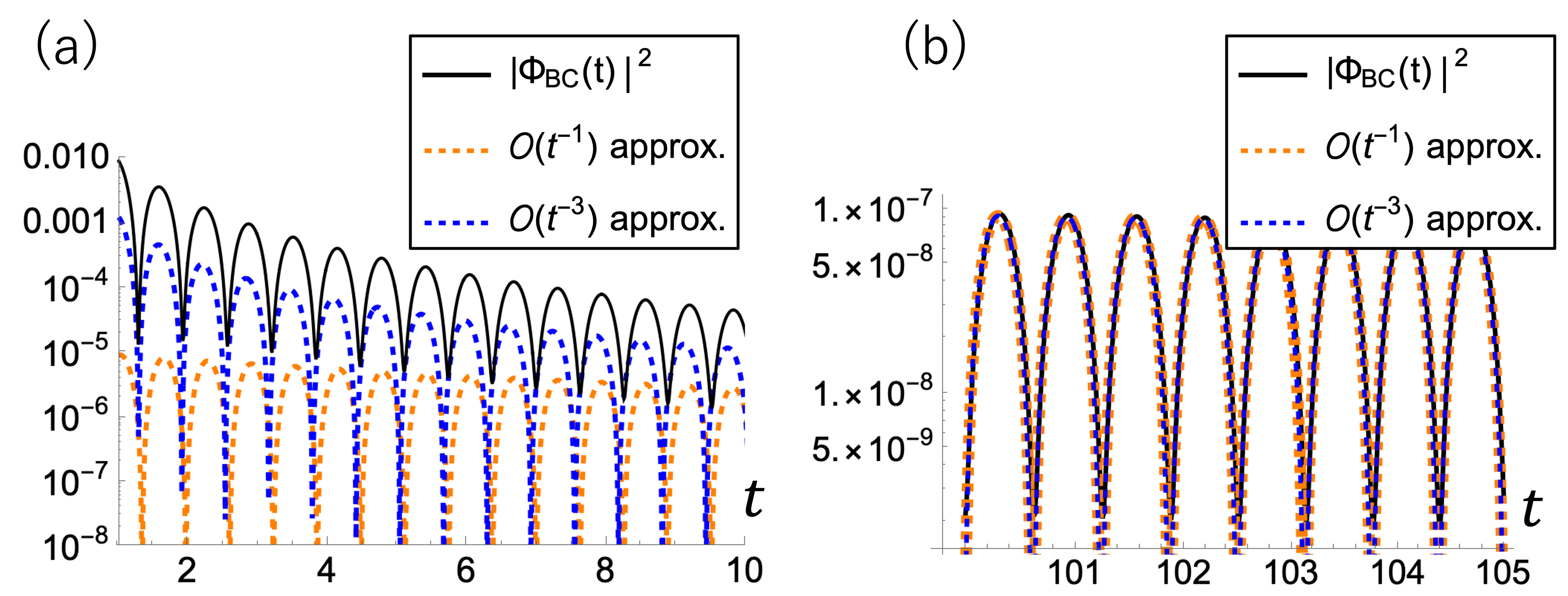}
        	\caption{Two panels show plots for $\omega_0 = 1$, $g = 1$, and $\Lambda = 5$. (a) A semi-logarithmic plot of the branch-cut contribution to the survival probability, $\left|\Phi_{\text{BC}}\right|^2$, given in Eq.~(\ref{Eq:L_finite_BC_solution}). The blue and orange dashed lines represent the asymptotic approximations up to orders $\mathcal{O}(t^{-1})$ and $\mathcal{O}(t^{-3})$, with the leading-order term given in Eq.~(\ref{Eq:m_zero_approx}). (b) The same plot as panel (a), but over the time range $t=100$ to $t=105$, where the full solution $\left|\Phi_{\text{BC}}\right|^2$ and the asymptotic approximations are in good agreement.}
        	\label{fig: L_finite_Prob_BC}
        \end{figure*}
        
        Let $F_n(z)$ denote the integrand of Eq.~(\ref{Eq:L_finite_formal_solution}) on the $n$th Riemann sheet $\mathcal{R}_n$, with the principal Riemann sheet denoted by $\mathcal{R}_0$. 
        Because of the function $\text{arctan}$, there are infinitely many Riemann sheets labeled by the integers $n\in \mathbb{Z}$, but we will only use $F_0 (z)$ and $F_1 (z)$ for our calculation. 
        The explicit form of $F_n(z)$ is given by
        \begin{align}
            F_n (z) =\left\{z+i \frac{\omega_0}{g^2}+ 4
        \left[\text{Arctan}\left( \frac{\Lambda / g^2}{z }\right)+ n \pi\right]\right\}^{-1},
        \end{align}
        where $\text{Arctan}(z)$ is a single-valued arctangent function defined on the principal Riemann sheet.
        
        Let $J_{\text{up}}$ denote the contribution of the upper branch cut in Fig.~9(a). It is given by
        \begin{align}
           J_{\text{up}}&= e^{\frac{i}{\hbar}\omega_0 t}\int_{-R+ i \Lambda/g^2}^{i \Lambda/g^2} \frac{dz}{2\pi i } ~  \left(F_0 (z) - F_1 (z)\right)e^{z\frac{g^2}{\hbar}t}\nonumber\\
            &= e^{\frac{i}{\hbar}\omega_0 t}\int_{-R}^{0} \frac{dw}{2\pi i } \mathbb{A}\left(w+i \Lambda / g^2 \right)e^{(w+i \Lambda /g^2)\frac{g^2}{\hbar}t} ,
        \end{align}
        where $\mathbb{A}(z) := F_0 (z)-F_1 (z)$ and we have performed the coordinate transformation $z = w + i \Lambda/g^2$. The contribution of the lower branch cut can be calculated similarly, and thus we find the full branch-cut contribution in the following form:
        \begin{align}\label{Eq:L_finite_BC_solution}
              \Phi_{\text{BC}}(t) &=  e^{\frac{i}{\hbar}\omega_0 t}\lim_{R\rightarrow\infty}\int_{-R}^{0} \frac{dw}{2\pi i } e^{w\frac{g^2}{\hbar}t}\nonumber\\
              &\times \big[\mathbb{A}\left(w+ i \Lambda / g^2\right)e^{i \frac{\Lambda}{\hbar} t} - \mathbb{A} \left(w-i \Lambda / g^2\right)e^{-i \frac{\Lambda}{\hbar} t}\big].
        \end{align}
        The above integral can be computed numerically and is shown as the solid lines in Fig.~\ref{fig: L_finite_Prob_BC}(a) and (b) for different time regimes.

        To analyze the decay profile towards the pole contributions, we now turn to the large-time asymptotic analysis of the branch-cut contribution~(\ref{Eq:L_finite_BC_solution}), outlined in Ref.~\cite{garmon2013amplification} again. 
        A similar procedure to the one for Eq.~(\ref{Eq:L_infty_approx}) this time produces
        \begin{align}\label{Eq:m_zero_approx}
             \Phi_{\text{BC}}(t) &=e^{\frac{i}{\hbar}\omega_0 t} \frac{-4   \sin \left(\Lambda  t/\hbar \right)}{\left(\omega_0 /g^2 - i 2\pi\right) \left(\omega_0 /g^2 - i 6\pi\right)} \left(\frac{\hbar}{g^2 t}\right)\nonumber\\
             &+ \mathcal{O}\left(\left(\frac{\hbar}{g^2 t}\right)^2\right).
        \end{align}
        Thus, we find that the asymptotic contribution from the branch cut again produces a power-law decay with an oscillation with frequency $\Lambda / \hbar$. 
        Notice that the exponent of the power-law decay differs from the one in the low-energy case in Eq.~(\ref{Eq:L_infty_approx}), where it was $t^{-3/2}$. 
        This implies that the decay in the case of a finite cutoff ($\Lambda < \infty$) with the massless Dirac particle ($m=0$) is faster than in the case of the infinite cutoff ($\Lambda \to \infty$) with the massive Dirac particle ($m \neq 0$) analyzed in Sec.~\ref{Sec:Low_energy_structure_and_Markovianity}.
        
        The long-time dynamics of the branch-cut contribution are illustrated in Fig.~\ref{fig: L_finite_Prob_BC}(a) and (b), where the full numerical result from Eq.~(\ref{Eq:L_finite_BC_solution}) is plotted alongside the asymptotic approximations up to orders $\mathcal{O}(t^{-1})$ and $\mathcal{O}(t^{-3})$. As shown in Fig.~\ref{fig: L_finite_Prob_BC}(b), the leading-order term given in Eq.~(\ref{Eq:m_zero_approx}) agrees well with the full solution for time $t \geq  100$, confirming the validity of the asymptotic expansion in this regime.

    \subsubsection{Survival probability: Short-time dynamics}\label{Sec: short time with finite cut-off}
        To analyze the short-time regime, we use the Dyson-series expansion given in Eq.~(\ref{Eq. Dyson series}). In the limit $m \rightarrow 0$, the memory kernel $\mathcal{K}(t)$, defined in Eq.~(\ref{Eq:Memory_kernel}), can be evaluated explicitly as
        \begin{align}
            \mathcal{K}(t) = - \frac{4 g^2}{\hbar}e^{\frac{i}{\hbar}\omega_0 t} \frac{\sin\left(\Lambda t / \hbar \right)}{t}.
        \end{align}
        With this, the first few terms of the exponent in the Dyson series are
        \begin{align}
             \int_0^{t} ds \int_0^{t-s}du ~\mathcal{K}(u) &= - 2 \frac{\Lambda}{g^2} \left(\frac{g^2}{\hbar}t\right)^2- i \frac{2}{3}\frac{\omega_0}{g^2}\frac{\Lambda}{g^2} \left(\frac{g^2}{\hbar}t\right)^3 \nonumber\\
             &+ \mathcal{O}\left(\left(\frac{g^2}{\hbar}t\right)^4\right).
        \end{align}  
        Thus, the short-time approximation of the solution to the non-Markovian dynamical equation~(\ref{Eq:Master equation simple form}) in the case of $m = 0 $ and $ \Lambda < \infty$ becomes
        \begin{align}
            \Phi(t)& \sim \exp \left[- 2 \frac{\Lambda}{g^2} \left(\frac{g^2}{\hbar}t\right)^2- i \frac{2}{3}\frac{\omega_0}{g^2}\frac{\Lambda}{g^2} \left(\frac{g^2}{\hbar}t\right)^3 \right],
        \end{align}
        as $g^2 t/ \hbar \rightarrow 0$.
        
        Using this approximation, the short-time behavior of the survival probability, $P(t) = \left|\Phi(t)\right|^2$, is given by
        \begin{align}\label{Eq: short time solution finite bound}
                P(t) &= 1 -4 \frac{\Lambda}{g^2} \left(\frac{g^2}{\hbar} t\right)^2 \\
                 & +\frac{\Lambda  \left[\Lambda  \left(72 g^2+\Lambda \right)+3 \omega _0^2\right]}{9 g^6}\left(\frac{g^2}{\hbar} t\right)^4 + \mathcal{O}\left(\left(\frac{g^2}{\hbar} t\right)^6\right).\nonumber
        \end{align}
        Thus, we find that the survival probability exhibits a quadratic decay in the short-time regime, indicating non-exponential behavior. The semigroup property is also checked in Appendix~\ref{Appdx: CP-divisibility} and shown to be violated, indicating non-Markovianity.
        
\begin{figure}[h]
        	\centering
        	\includegraphics[width=0.4\textwidth]{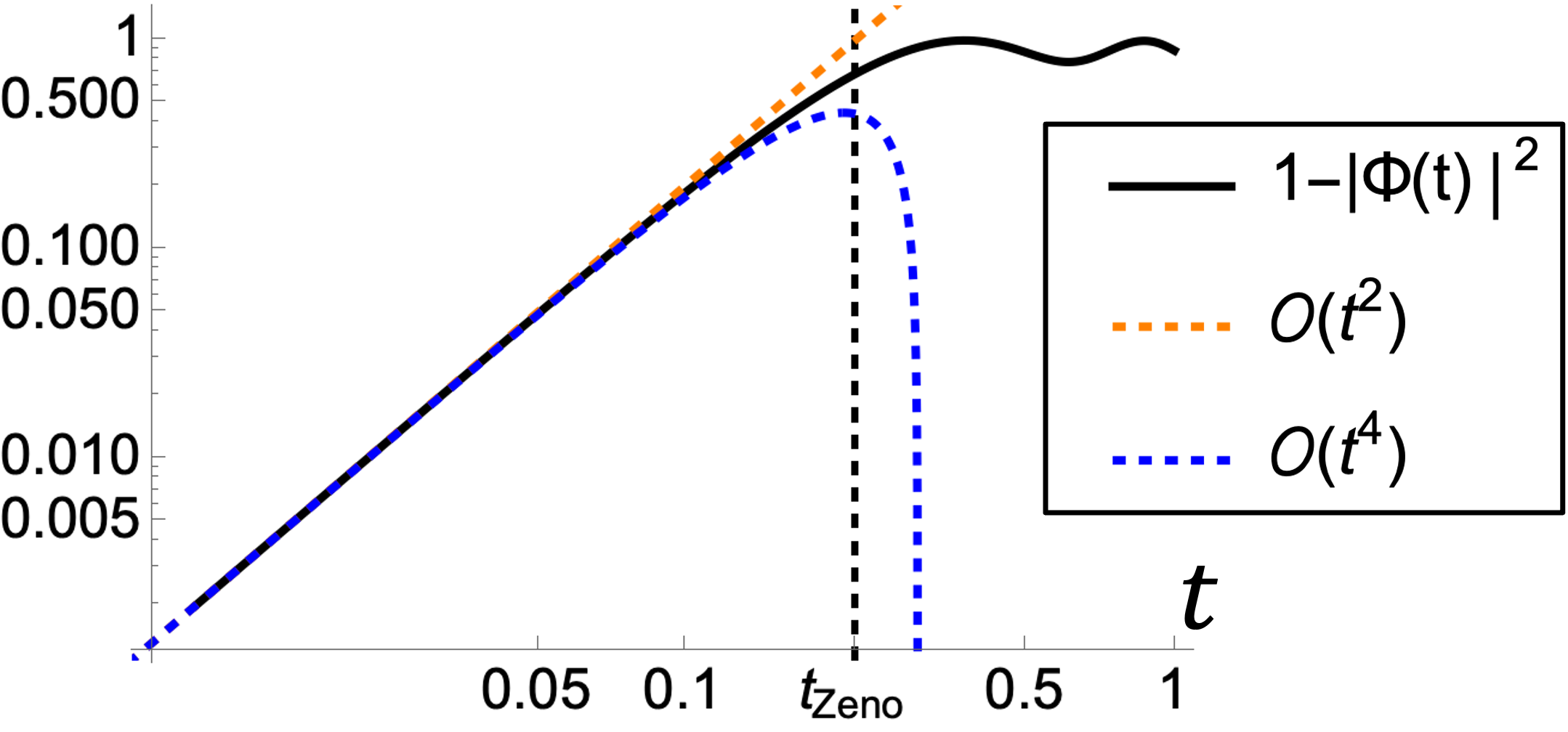}
        	\caption{Logarithmic plot of the survival probability $1-\left|\Phi(t)\right|^2$, where $\Phi(t) = \text{Pole}(t) + \Phi_{\text{BC}}(t)$, with $\text{Pole}(t)$ given in Eq.~(\ref{Eq: Pole contribution}) and $\Phi_{\text{BC}}(t)$ given in Eq.~(\ref{Eq:L_finite_BC_solution}). The plot is compared with the asymptotic expansions up to $\mathcal{O}(t^2)$ and $\mathcal{O}(t^4)$ from Eq.~(\ref{Eq: short time solution finite bound}). The vertical line indicates the Zeno time defined in Eq.~(\ref{Eq: Zeno time}). Parameter values are $\omega_0 = 1$, $g = 1$, and $\Lambda = 5$.}
        	\label{fig: LFiniteShortSP}
        \end{figure}

        The short-time dynamics of the survival probability $P(t)$ is depicted in Fig.~\ref{fig: LFiniteShortSP}, where we use a logarithmic plot to display $1 - \left|\Phi(t)\right|^2$, with $\Phi(t) = \text{Pole}(t) + \Phi_{\text{BC}}(t)$ given by Eqs.~(\ref{Eq: Pole contribution}) and~(\ref{Eq:L_finite_BC_solution}), respectively. Two dashed lines represent the short-time dynamics obtained from Eq.~(\ref{Eq: short time solution finite bound}) up to orders $\mathcal{O}(t^2)$ and $\mathcal{O}(t^4)$. Notably, the plot of $1 - \left|\Phi(t)\right|^2$ aligns with the quadratic-order approximation, indicating that the quantum Zeno effect (see Appendix~\ref{Appdx: Quantum Zeno Effect}) is present, with the Zeno time obtained from the leading-order terms in Eq.~(\ref{Eq: short time solution finite bound}) as in $1-4 (\Lambda / g^2) (g^2 t_{\text{zeno}}/ \hbar)^2 = 0$, producing 
        \begin{equation}\label{Eq: Zeno time}
            t_\text{Zeno} = \frac{\hbar}{2 g \sqrt{\Lambda}},
        \end{equation}
        which sets the scale of the time interval for the continuous measurement to induce the quantum Zeno effect. The Zeno time is indicated by a vertical dashed line in Fig.~\ref{fig: LFiniteShortSP}; we observe good agreement of the $\mathcal{O}(t^2)$ and $\mathcal{O}(t^4)$ approximations up to the Zeno time $t_{\text{Zeno}}$.
        
        Finally, summarizing our analysis of the survival probability in the short- and long-time regimes in the case of $m=0$ and $\Lambda < \infty$, we obtain the following asymptotic expressions for the probability amplitude:
        \begin{widetext}
        \begin{subequations}\label{Eq: Asym-exact 2}
            \begin{align}
                \Phi(t) &\sim \text{exp}\left[ - 2 \frac{\Lambda}{g^2} \left(\frac{g^2}{\hbar} t \right)^2\right]  ~ \text{ as }~ \frac{g^2}{\hbar} t\rightarrow 0 ,\label{Eq: Asym-exact short 2}\\
                \Phi(t) & \sim \Phi_{ix_1} (t) + \Phi_{-ix_2} (t) -\frac{4   \sin \left(\Lambda  t/\hbar \right)e^{\frac{i}{\hbar}\omega_0 t}}{\left(\omega_0 /g^2 - i 2\pi\right) \left(\omega_0 /g^2 - i 6\pi\right)} \left(\frac{\hbar}{g^2 t}\right)~\text{ as }~ \frac{g^2}{\hbar} t\rightarrow \infty,\label{Eq: Asym-exact long 2}
            \end{align}
        \end{subequations}
        \end{widetext}
        where $ \Phi_{ix_1} (t)+ \Phi_{-ix_2}(t)$ is the pole contribution given in Eq.~(\ref{Eq: Pole contribution}). The semigroup property in both short- and long-time regimes is violated (see appendix~\ref{Appdx: CP-divisibility}), indicating the non-Markovianity.

    \subsubsection{Wave function of environmental states}\label{sec: long time with finite cut-off}
         The wave function in the environment can be evaluated by taking the limit $m\rightarrow 0$ of Eq.~(\ref{Eq: Wave function}):
        \begin{align}
            \psi(t,x)&= \frac{1}{i} \sqrt{\frac{2g^2 \pi}{\hbar}} \lim_{R \rightarrow \infty} \int_{-iR + \sigma}^{iR + \sigma} \frac{dz}{2\pi i} e^{\frac{z}{\hbar} t} F(z) \nonumber\\
            &\quad \times \int_{-\Lambda}^\Lambda dp~ \frac{z \cos\left( \frac{p x}{\hbar} \right) }{z^2 + p^2}, \label{Eq: Wave function of finite cut-off}
        \end{align}
        The momentum integral does not yield a closed-form expression in terms of elementary functions. Therefore, an analysis analogous to the one in  Sec.~\ref{Sec:Low_energy_structure_and_Markovianity}, where the wave function was explicitly decomposed into bound-state and branch-cut contributions, is not possible. Since the main focus is to see how the above wave function approaches to the time-evolving resonant state observed in Sec .~\ref {sec: L infinite and m zero}, a detailed analysis of this wave function is left for future work.

\begin{figure}[ht]
        	\centering
        	\includegraphics[width=0.35\textwidth]{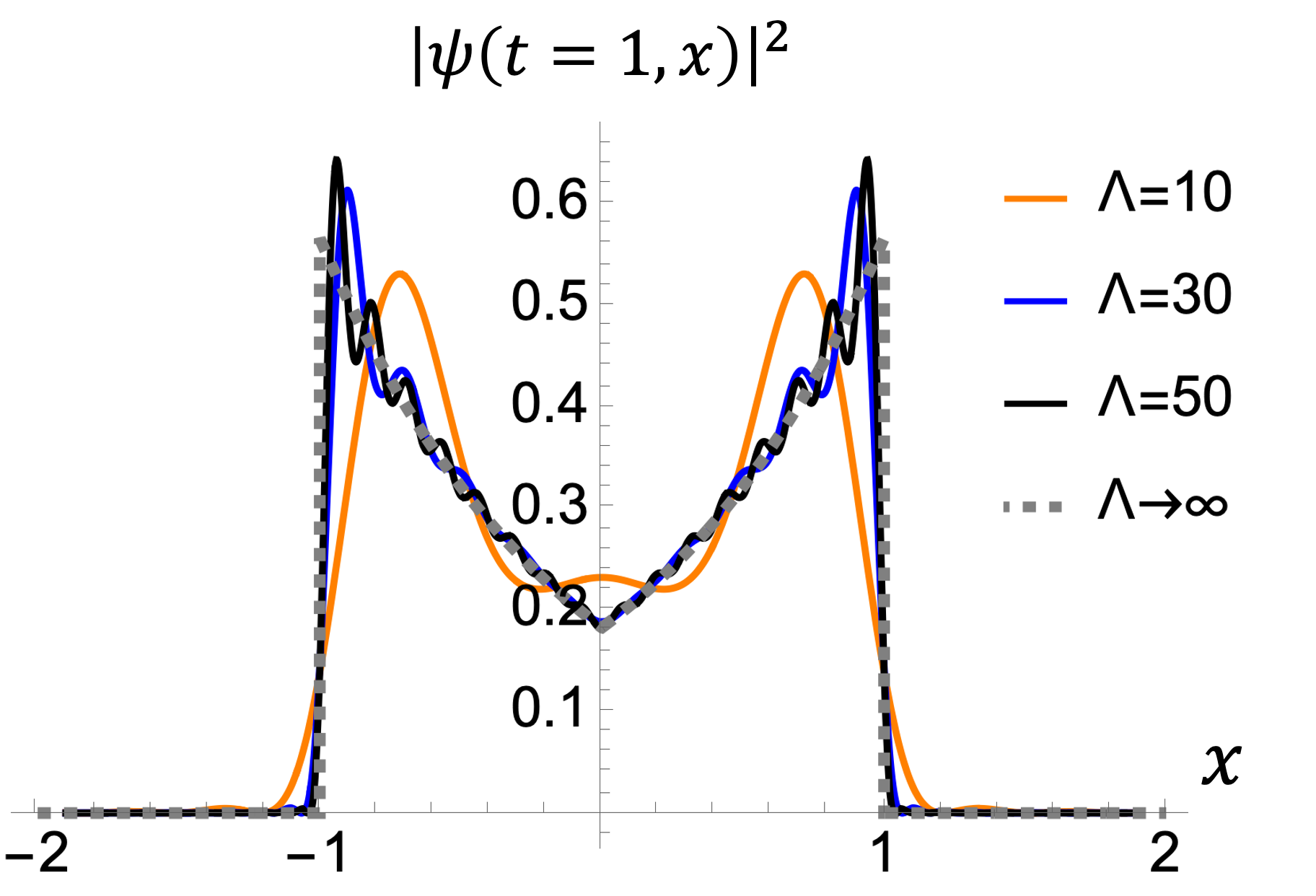}
        	\caption{The probability density of the time-evolving state described by Eq.~(\ref{Eq: Wave function of finite cut-off}), plotted for $\omega_0=1$, $m = 0$, and $g=0.3$, with varying cutoff $\Lambda=10$, $\Lambda =30$ and $\Lambda = 50$. $\Lambda\rightarrow \infty$ shows the time-evolving resonant state (\ref{Eq: Exp wave function}).}
        	\label{fig: L finite scattering state}
        \end{figure}

        To examine how the wave function~(\ref{Eq: Wave function of finite cut-off}) behaves under a finite cutoff \(\Lambda < \infty\), we numerically plot the probability density of the wave function in the limit \(m \rightarrow 0\).
        Figure~\ref{fig: L finite scattering state} shows the wave function for \(\omega_0 = 1\), \(m = 0\), \(g = 0.3\), and \(t = 1\), for three values of the cutoff: \(\Lambda = 10\), \(30\), and \(50\). 
        As shown in Fig.~\ref{fig: L finite scattering state}, the time-evolving state approaches the time-evolving resonant state observed in Sec.~\ref{sec: L infinite and m zero}, plotted as dashed lines. Similar to the previous subsection~\ref{sec: wave function of massive no-cutoff case}, we can interpret the time-evolving states corresponding to finite cutoff \(\Lambda = 10\), \(30\), and \(50\) as an approximate time-evolving resonant state.  
        The oscillations near the causal point \(t = |x|\) become increasingly vibrant as the cutoff increases, \(\Lambda \rightarrow \infty\). 
        This Gibbs phenomenon was observed in Sec.~\ref{sec: wave function of massive no-cutoff case} too. With the finite cutoff, the wave function spreads slightly beyond the causality boundary $t = |x|$, but it has been shown that this does not violate causality in the finite-cutoff case~\cite{petrosky2001space}.

\subsection{Low- and high-energy structure: $m\not=0$ and $\Lambda< \infty$}\label{sec: full analysis}
        \begin{figure*}[t]
        	\centering
        	\includegraphics[width=0.7\textwidth]{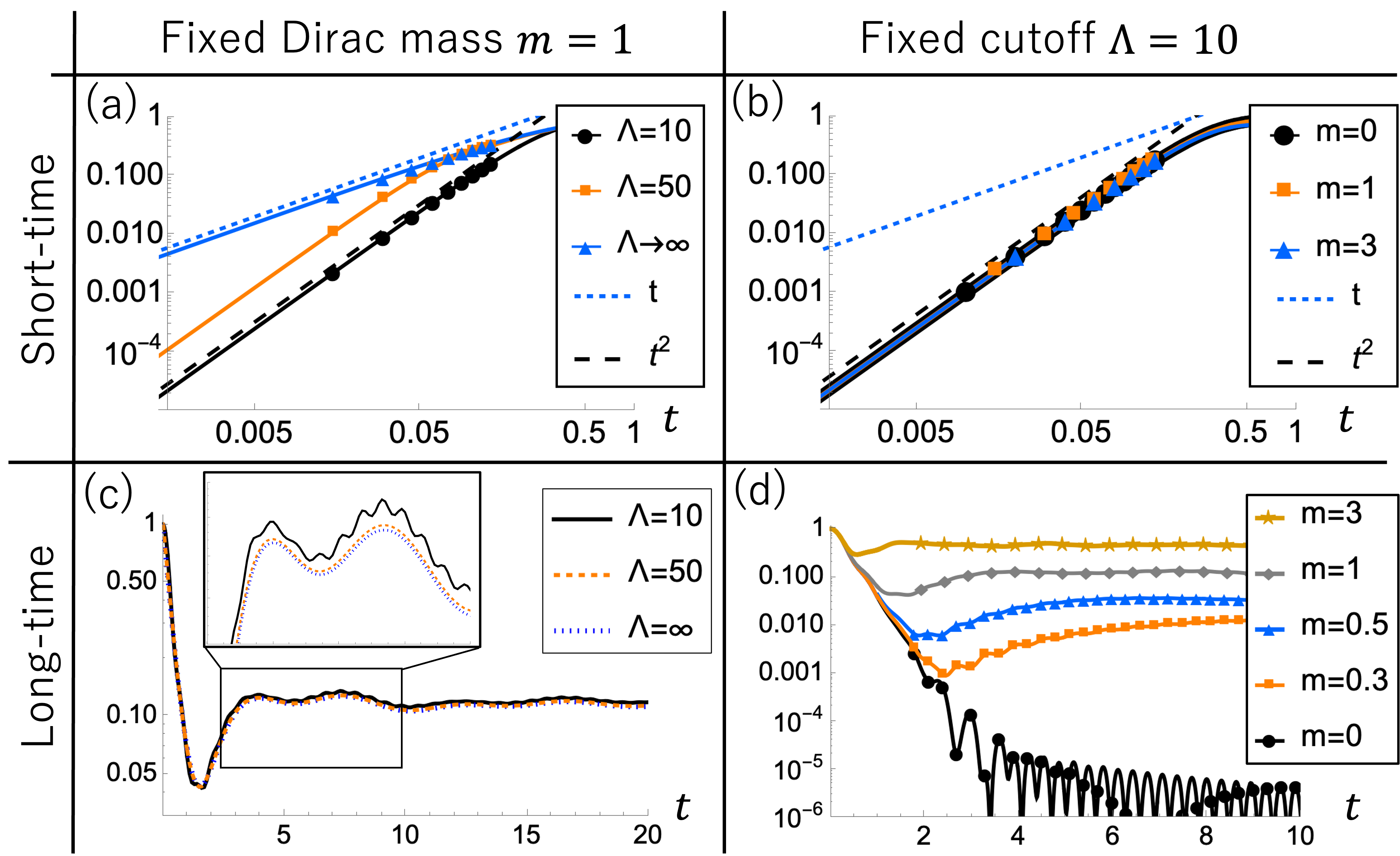}
        	\caption{Numerical solutions of the inverse Laplace transform in Eq.~(\ref{Eq:full_solution}) for short- and long-time regimes with parameter values $\omega_0 = 1$, $g = 0.5$, and varying values of the Dirac gap $2m$ and cutoff $\Lambda$.  
            (a) Logarithmic plot of $1 - \left|\Phi\right|^2$ for different values of the cutoff $\Lambda$ with fixed Dirac mass $m=1$. As $\Lambda \to \infty$, the decay transitions from a quadratic one to a linear one.  
            (b) Logarithmic plot of $1 - \left|\Phi\right|^2$ for different values of $m$ with a fixed cutoff $\Lambda$. The short-time dynamics remain nearly unchanged, consistent with the findings of Sec.~\ref{Sec: Bound state}, where $m$ does not appear in the short-time expansion.  
            (c) Semi-logarithmic plot of the survival probability for a fixed Dirac mass $m=1$ with varying $\Lambda$. Unlike panel (a), varying $\Lambda$ from $10$ to $\infty$ results in only minor changes to the dynamics.  
            (d) Semi-logarithmic plot of the survival probability for a fixed cutoff $\Lambda$ with varying Dirac gap $2m$. Unlike panel (c), varying $m$ visibly alters the asymptotic behavior, suggesting that the bound-state amplitude is primarily determined by $m$ rather than $\Lambda$.}
        	\label{fig: full plot}
        \end{figure*}
        
        In this section, we numerically consider the case $m \neq 0$ and $\Lambda < \infty$ for completeness. Exact analyses of this case is challenging, as it requires directly treating poles and the branch-cut structure of the integrand in Eq.~(\ref{Eq:full_solution}), which involves a transcendental equation with a square root. Therefore, we leave the analytic treatment of this case for future work.  
        
        We plot the numerical solution of the inverse Laplace transform of Eq.~(\ref{Eq:full_solution}) in Fig.~\ref{fig: full plot}, showing the short- and long-time regimes in panels (a), (b), (c), and (d), respectively. All plots in Fig.~\ref{fig: full plot} correspond to the parameter values $\omega_0 = 1$, $g = 0.5$, with varying values of the Dirac gap $2m$ and the cutoff $\Lambda$.  
        
        Panel (a) shows a logarithmic plot of the quantity $1 - \left|\Phi\right|^2$, where $\left|\Phi\right|^2$ is given in Eq.~(\ref{Eq:full_solution}), for different values of the cutoff $\Lambda$ with a fixed Dirac gap of $2m=2$. As the cutoff parameter approaches infinity, the survival probability transitions from a quadratic decay to a linear decay. This behavior is consistent with our observations in Secs.~\ref{Sec: Bound state} and~\ref{Sec: short time with finite cut-off}.  
        
        Panel (b) also shows a logarithmic plot of $1 - \left|\Phi\right|^2$, but for varying the Dirac gap $2m$ with a fixed cutoff $\Lambda$. In contrast to panel (a), we observe that the short-time dynamics remains nearly unchanged for different values of $m$. This behavior is consistent with our findings in Sec.~\ref{Sec: Bound state}, where we showed that the effect of having a finite Dirac gap $2m$ does not appear in the short-time dynamics.  
        
        Panel (c) shows a semi-logarithmic plot of the survival probability for a fixed Dirac gap of $2m=2$ with varying values of the cutoff $\Lambda$. In contrast to panel (a), we observe only a small change in the dynamics as we vary the cutoff from $\Lambda = 10$ to $\Lambda = \infty$. The behavior in panel (c) was not observed in Sec.~\ref{sec: short time with finite cut-off}, where we found Rabi oscillations with an $\mathcal{O}(t^{-2})$ decay. We also find that the survival probability saturates to a non-zero value at late time, suggesting the presence of a bound state in the spectrum. Confirming the existence of such a bound state requires an analysis of the pole and branch-cut structure of the resolvent~(\ref{Eq:resolvent}). This is numerically challenging because the resolvent exhibits a nontrivial multivalued structure. As a result, its complex branch-cut topology makes it difficult to identify which poles lie on the principal Riemann sheet and contribute to the dynamics. We defer this for future work.

        Panel (d) shows a semi-logarithmic plot of the survival probability for a fixed cutoff $\Lambda$ with varying values of the Dirac gap $2m$. In contrast to panel (c), we observe a visible change in the dynamical behavior for different values of $m$, namely, once the gap opens, the survival probability asymptotes to a finite, nonzero value, and this asymptotic value increases with $m$. This behavior suggests the existence of a bound state, but, as discussed above, it is challenging to confirm conclusively.

        In conclusion, Fig.~\ref{fig: full plot} (a)--(d) demonstrate that the short-time behavior of the survival probability is dominated by the cutoff $\Lambda$, while the long-time behavior is dominated by the Dirac gap $2m$.
        These findings are summarized in Table~\ref{tab:results}.
        In the next section, we propose experimental setups to test our theoretical findings.

\section{Experimental proposal}\label{sec: experiment}
    % What we might see in the experiment is the appearance of 
    \subsection{Possible experimental setups}
    To test our theoretical results experimentally, the key is to realize a one-dimensional dispersion relation with a controllable gap and cutoff structure that approximately mimics the Dirac dispersion relation, $\omega_p = \pm\sqrt{m^2 + p^2}$, in momentum space as the environment. This can be achieved in various experimental setups, such as graphene nanoribbons~\cite{matthes2014influence}, on-surface synthesis of Su-Suchriffer-Heeger chains~\cite{jalochowski2024implementation,grill2020covalent}, topological waveguide QED~\cite{bello2019unconventional}, and optical waveguide arrays~\cite{crespi2019experimental}. Below, we briefly elaborate on three possible experimental implementations.  
    
    \paragraph{Optical waveguide array}  
    The optical setup illustrated in Fig.~\ref{fig: Experimental set up}(a) consists of a single optical waveguide coupled to an optical array in the SSH configuration. Coherent laser light is injected into the single waveguide and propagates along the $l$-direction, with photons tunneling into adjacent waveguides. The fraction of photons remaining in the initial waveguide corresponds to the survival probability. The photon dynamics in waveguides are governed by the paraxial approximation of Maxwell's equations called the Helmholz equation, which, under appropriate conditions, are mathematically equivalent to the Schr\"{o}dinger equation, where the waveguide depth $l$ plays the role of time. Formally, the mapping $t/\hbar \leftrightarrow l$ connects the two equations. A similar experiment has already been performed with equal hopping strengths $t_1 = t_2 = t$ of the SSH configuration~\cite{crespi2019experimental}, where the observed survival probability matches the theoretical predictions discussed in Sec.~\ref{Sec:High_energy_structure_and_Markovianity} of the present work.    
    
    \paragraph{Graphene nanoribbon}  
    The nanoribbon setup consists of a two-level system coupled to a graphene nanoribbon, where electrons effectively behave as the Dirac particles with a dispersion relation that can exhibit both gapped and gapless phases~\cite{matthes2014influence}. To our knowledge, no experiment has yet been conducted with an external two-level system attached to a nanoribbon. However, a proposal exists for realizing an intrinsic two-level structure within a nanoribbon~\cite{braun2011spin}, which could serve as the system in our framework.  
    
    \paragraph{Topological waveguide QED}  
    Another possible platform is a topological waveguide quantum electrodynamics (QED)~\cite{bello2019unconventional}, where a single or multiple qubits are attached to a photonic analogue of the SSH model. An experimental setup proposed in Ref.~\cite{kim2021quantum} shows a qubit attached to a photonic lattice, consisting of quantum inductor-capacitor resonators coupled via capacitors with alternating capacitance. This setup realizes the photonic version of the SSH model attached to the qubit.  
    
    Among the three proposed setups, the optical waveguide is the only experiment in which detailed measurement of the survival probability has been performed in both short- and long-time regimes~\cite{crespi2019experimental}. This allows us to use typical parameter ranges from the experiment to determine whether our results fall within an experimentally feasible regime. We discuss this point further in the next subsection~\ref{Sec: optical experiment}.

    \subsection{Optical experiment}\label{Sec: optical experiment}
    
\begin{figure*}[ht]
            	\centering
            	\includegraphics[width=0.7\textwidth]{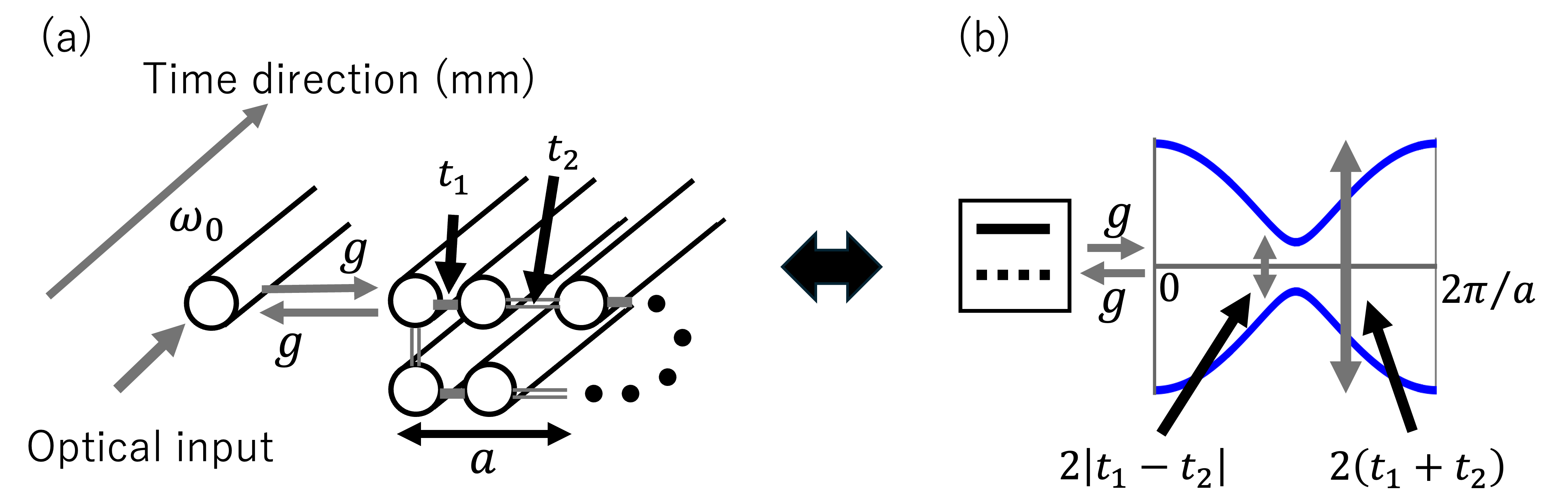}
            	\caption{Panel (a) shows an experimental setup of a single waveguide with propagation constant $\omega_0$, coupled to a waveguide array in the SSH configuration with periodic boundary conditions. Panel (b) shows the model in momentum space, where the two-level system is coupled to the cosine band dispersion relation.}
            	\label{fig: Experimental set up}
            \end{figure*}

            We focus on an optical experiment, as a recent study has demonstrated that both short- and long-time regimes can be investigated within the same experimental setup~\cite{crespi2019experimental}. In particular, we propose a new experimental setup by using an optical waveguide array in the SSH configuration, shown schematically in Fig.~\ref{fig: Experimental set up}(a). 

            To express the corresponding Hamiltonian, we denote each optical mode by $\ket{n}_A$ and $\ket{n}_B$ with $n=0,1,2,\dots $ where $A$ and $B$ indicates the alternating SSH sites. The system site is denoted as $\ket{0}_S$. The Hamiltonian for the setup in Fig.~\ref{fig: Experimental set up}(a) is given by 
            \begin{align}\label{Eq: Optical Hamiltonian}
                H = H_0 + H_{\text{SSH}} + H_{\text{Int}},
            \end{align}
            where 
            \begin{align}\label{Eq: qubit-SSH model}
                &H_0 = \omega_0 \ket{0}_S \prescript{}{S}{\bra{0}} , \quad H_\text{Int} = g (\ket{0}_S \prescript{}{A}{\bra{0}} + \ket{0}_A\prescript{}{S}{\bra{0}}), \\
                &H_\text{SSH} = t_1 \sum_{n=0}^{N-1}  \Big(\ket{n}_A\prescript{}{B}{\bra{n} }+ \ket{n}_B \prescript{}{A}{\bra{n}}\Big) \nonumber\\
                &+t_2\sum_{n=0}^{N-1} \Big(\ket{n+1}_A \prescript{}{B}{\bra{n}} + \ket{n}_B \prescript{}{A}{\bra{n+1}} \Big),
            \end{align}
            with the periodic boundary condition $\ket{N}_A = \ket{0}_A$ to simplify the theoretical calculation. The parameter $\omega_0$ is the propagation constant of the system waveguide, which is coupled to the SSH array with strength $g$. The SSH array consists of coupled optical waveguides with alternating nearest-neighbor couplings $t_1$ and $t_2$.

            The dynamics of the optical mode traveling through the waveguide array can be approximately described by the Helmholtz equation, in which time is replaced by the waveguide depth $l$. This equation is formally equivalent to the Schr\"{o}dinger equation: $i\partial_l \ket{\phi (l)} = H \ket{\phi (l)}$. With the SSH Hamiltonian diagonalized in momentum space, as illustrated in Fig.~\ref{fig: Experimental set up}(b), the procedure of Sec.~\ref{Sec: Master equation} carries over directly and yields the non-Markovian dynamical equation for the system's probability amplitude $\Phi(l)=e^{i\omega_0 l} \braket{0}{\phi(l)} $: 
            \begin{align}\label{Eq: Optical master equation}
                \frac{d \Phi(l)}{dl}&= - \frac{ag^2}{2\pi}\int_0^l ds e^{i \omega_0 (l-s) }\Phi (s)\nonumber \\
                &\times \int_{-\pi / a}^{\pi / a} dk \cos \left[\omega_k (l-s)\right],
            \end{align}
            where $a$ is the lattice spacing in the SSH chain. The dispersion relation for the optical setup is 
            \begin{align}
                \omega_k = \sqrt{t_1^2 + t_2^2 - 2 t_1 t_2 \cos(k a )}.
            \end{align}
            Note that the coupling strengths $t_1$ and $t_2$ have dimensions of inverse length, and are thus inversely proportional to the spacing between waveguides~\cite{crespi2019experimental}. Taking the Laplace transform of the above equation would lead to the formal solution given by the Bromwich integral:
            \begin{align}
                \Phi(l) &= \frac{i}{2\pi}e^{i\omega_0 l} \lim_{R\rightarrow \infty}\int_{-i R + \sigma}^{i R + \sigma} \frac{dz}{2\pi i} e^{z l}\nonumber \\
                &\times \left(z+i \omega_0+ag^2\frac{z}{\sqrt{z^2+m^2}\sqrt{z^2 + L^2}} \right)^{-1} \Phi(0),
            \end{align}
            where we have denoted $m =\left|t_1 -t_2\right| $ and $L = t_1 + t_2$. Note that \(L\) corresponds to the spectral bound \(\sqrt{\Lambda^{2}+m^{2}}\) of the Dirac Hamiltonian in Sec.~\ref{Sec: Master equation}. Therefore, taking the limit $L\rightarrow \infty$ reproduces the solution~(\ref{Eq:Simpler_solution}) in Sec.~\ref{Sec: Master equation}.  Furthermore, we choose the initial boundary condition on the optical array to be $\Phi(0)=1$, \textit{i.e.}, the initial state is $\ket{\phi(0)}=\ket{0}$ with no occupation of the SSH modes (the array in vacuum).
            
\begin{figure*}[ht]
            	\centering
            	\includegraphics[width=\textwidth]{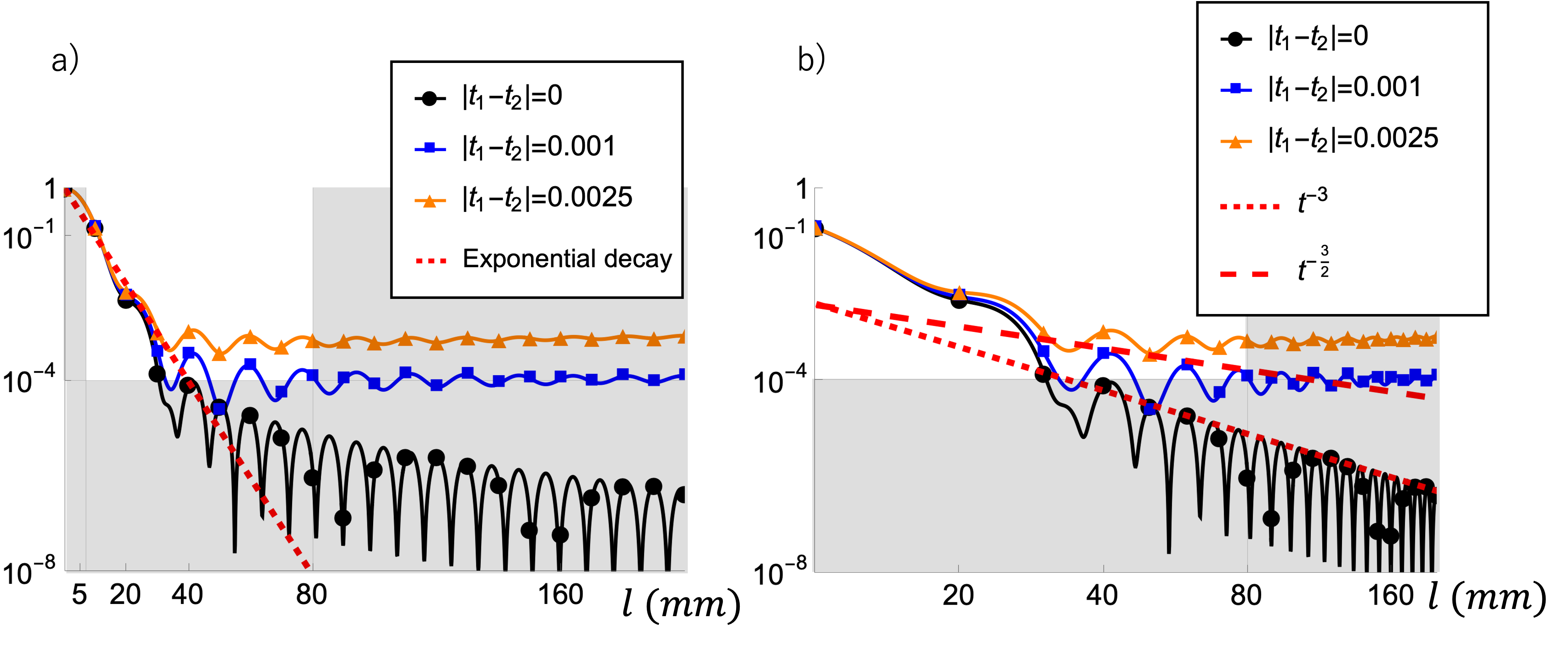}
            	\caption{(a) Theoretical prediction of the survival probability in the optical setup illustrated in Fig.~\ref{fig: Experimental set up}. Solid lines denote the survival probability for \( \omega_0 = 0.01\,\mathrm{mm}^{-1} \), \( g = 0.136\,\mathrm{mm}^{-1} \), and \( t_1 + t_2 = 0.36\,\mathrm{mm}^{-1} \), with varying values of the SSH coupling strength difference \( |t_1 - t_2| = 0, 0.001, 0.0025 \) represented by different markers. Dashed line denotes a reference curve for exponential decay.
(b) Logarithmic plot of the numerical data presented in panel (a). The dashed lines indicate reference decay rates corresponding to \( t^{-3} \) and \( t^{-3/2} \), respectively.}
            	\label{fig: optical survival probability 1}
            \end{figure*}

            As a demonstration, we examine one of our main results from Sec.~\ref{Sec:Low_energy_structure_and_Markovianity}, namely the crossover from non-exponential to exponential decay as the gap is closed. Since we cannot take the limit $L\rightarrow \infty$ in an experiment, we choose $t_1 $ and $t_2$ using the largest experimentally accessible values reported in Ref.~\cite{crespi2019experimental}, namely $t_1 = t_2 = 0.18\,\mathrm{mm}^{-1}$. For other parameters, we take the similar parameter values from the Ref.~\cite{crespi2019experimental} as $g=0.136\,\mathrm{mm}^{-1}$, and $ \omega_0 = 0.01\,\mathrm{mm}^{-1}$.
            The resulting decay is numerically plotted by diagonalizing the tight-binding Hamiltonian~(\ref{Eq: qubit-SSH model}) in a semi-logarithmic plot in Fig.~\ref{fig: optical survival probability 1}(a). Its logarithmic plot is shown in Fig.~\ref{fig: optical survival probability 1}(b).
            In the experimentally visible region (the unshaded area in Fig.~\ref{fig: optical survival probability 1}(a)), we observe an exponential decay up to $l=40$ as shown in Fig.~\ref{fig: optical survival probability 1}(a) for $\left|t_1-t_2\right|=0$, followed by a crossover to a power-law decay at large $l=40$. The nature of the power-law decay outside the experimentally visible region is more clearly seen in Fig.~\ref{fig: optical survival probability 1}(b).  Because this power-law regime lies outside the experimentally visible region, we expect that an experiment would observe only the portion of the exponential decay in the gapless case~$\left|t_1 - t_2\right|=0$. 

            Next, we gradually open the gap $\left| t_1-t_2\right |$, and test whether non-exponential decay can enter the experimentally visible region. As shown in Fig.~\ref{fig: optical survival probability 1}(a), for $\left| t_1-t_2\right | = 0.001 , 0.0025$, the survival probability shows an exponential decay up to $l=30$, followed by a power-law decay and then asymptotes to a finite value due to the existence of the bound state. The power-law decay is more clearly seen in the logarithmic plot in Fig.~\ref{fig: optical survival probability 1}(b). For $\left| t_1-t_2\right | = 0.001$ in the region between $30<l<80$, the decay profile aligns with $t^{-3/2}$. Because this power-law regime occurs within the unshaded region, the crossover from exponential to non-exponential decay induced by the gap should be observable experimentally.

\section{Conclusion}\label{Sec: conclusion}
We have examined the dynamics of the survival probability of a qubit coupled to a Dirac-particle bath, with particular attention to the interplay between short- and long-time behavior and the bath's low- and high-energy spectral features. We have also conducted an asymptotic analysis on the probability amplitudes of the qubit’s excited state across three distinct parameter regimes: \( (\Lambda \rightarrow \infty, m = 0) \), \( (\Lambda \rightarrow \infty, m \neq 0) \), and \( (\Lambda < \infty, m = 0) \). In each case, we have derived the asymptotically exact forms, as presented in Eqs.~(\ref{Eq: Surv. Prob. 1}), (\ref{Eq: Asym-exact 1}), and (\ref{Eq: Asym-exact 2}), respectively. These results enabled the identification of the exact power-law decay exponents. This was used to demonstrate that the semigroup property is preserved solely in the regime \( (\Lambda \rightarrow \infty, m = 0) \), whereas it breaks down in both \( (\Lambda \rightarrow \infty, m \neq 0) \) and \( (\Lambda < \infty, m = 0) \), thereby indicating the non-Markovian dynamics.

We have also analyzed the wave function emitted by the qubit into the Dirac-particle bath. We have shown that, in the Markovian limit, the wave function in the environment asymptotically converges to the time-evolving resonant state~\cite{HatanoNishino}, for which the normalizability is ensured by the finite spatial support imposed by causality. 

We have proposed an experimental scheme for the realization of the model within an optical waveguide array in Sec.~\ref{sec: experiment}. The feasibility of experimental implementation was examined via numerical analysis. We have shown that, under experimentally accessible parameters, the transition from exponential to non-exponential decay induced by the opening of the Dirac gap occurs within a viable experimental window.

In this study, we have explored the relationship between spectral structures and the decay profiles. Future work will aim to generalize this interplay to broader classes of models, assess the universality of the observed relationships, and investigate potential applications for characterizing non-Markovianity. Given the scope of open quantum systems, applications for diverse fields, such as quantum information, quantum optics, quantum chemistry, and nuclear physics, are anticipated. Notably, non-Markovian dynamics have been reported to enhance various quantum phenomena, including dynamical decoupling~\cite{addis2015dynamical,berk2023extracting}, quantum control~\cite{bylicka2013non,mirkin2019information,erez2008thermodynamic,reich2015exploiting}, quantum heat engines~\cite{thomas2018thermodynamics,bhattacharya2020thermodynamic}, quantum teleportation~\cite{laine2014nonlocal}, quantum error mitigation~\cite{endo2025non,hakoshima2021relationship}, and decoherence control~\cite{bylicka2014non}. Applications to these domains represent promising directions for future research.

\section*{Acknowledgments}
T.T. acknowledges financial support from JSPS Grant Numbers 22J01230, and R7 (2025) Young Researchers Support Project, Faculty of Science, KYUSHU UNIVERSITY. N.H. acknowledges financial support from JSPS KAKENHI Grant Numbers JP24K00545, and JP23K22411. The authors are grateful to Joshua Feinberg, Savannah Garmon, Eva-Maria Graefe, Sang Soon Oh, and Tomio Petrosky for fruitful discussions and helpful suggestions.

\appendix
\begin{widetext}
\section{Laplace transform}\label{Appdx: Laplace}

We here derive Eq.~(\ref{Eq:full_solution}) from the non-Markovian dynamical equation (\ref{Eq:Master equation simple form}). Instead of evaluating the memory kernel $\mathcal{K}$ in Eq.~(\ref{Eq:Memory_kernel}) directly, we perform the Laplace transform $\text{Lap}\left[\cdot\right]$ on Eq.~(\ref{Eq:Master equation simple form}), obtaining
\begin{align}\label{Eq: Laplace transformed master}
    z \overline{\Phi} (z) - \Phi (0) = \overline{\Phi} (z) \text{Lap}\left[\mathcal{K}(t)\right](z),
\end{align}
where $\overline{\Phi} (z) := \text{Lap}\left[\Phi (t)\right](z)$ and the Laplace transform is defined as follows:
\begin{align}
     \text{Lap}\left[\Phi (t)\right](z) := \int_0^{\infty} \Phi(t) e^{-zt}dt.\label{Eq: defintion of Laplace transform}
\end{align}
Note that the convolution in Eq.~(\ref{Eq:Master equation simple form}) is transformed to a product in Eq.~(\ref{Eq: Laplace transformed master}).

Let us examine the Laplace transform of the memory kernel in detail:
\begin{align}
    \text{Lap}\left[\mathcal{K}(t)\right](z) &= -\frac{2g^2}{\hbar^2} \text{Lap}\left[e^{\frac{i}{\hbar}\omega_0 t} \int_{-\Lambda}^\Lambda dp~ \cos\left(\omega_p \frac{t}{\hbar}\right)\right](z) \nonumber \\
    &= -\frac{2g^2}{\hbar^2} \text{Lap}\left[\int_{-\Lambda}^\Lambda dp~\cos\left(\omega_p \frac{t}{\hbar}\right)\right]\left(z - \frac{i}{\hbar} \omega_0 \right) \nonumber\\
    &= -\frac{2g^2}{\hbar^2} \int_{-\Lambda}^\Lambda dp~ \text{Lap}\left[ \cos\left(\omega_p \frac{t}{\hbar}\right)\right]\left(z - \frac{i}{\hbar} \omega_0 \right). \nonumber
\end{align}
In the second equality, we used the standard property of the Laplace transform, $\text{Lap} \left[e^{at} f(t)\right](z) = \text{Lap} \left[f(t)\right](z-a) $. 
In the third equality, we exchange the order of the integral with respect to $p$ and the Laplace transform.
The Laplace transform of the cosine function is given by
\begin{align}
    \text{Lap}\left[ \cos\left(\omega_p \frac{t}{\hbar}\right)\right](\Tilde{z}) = \frac{\Tilde{z} \hbar^2}{\Tilde{z}^2 \hbar^2 + \omega_p^2} = \frac{\Tilde{z} \hbar^2}{\Tilde{z}^2 \hbar^2 + m^2 + p^2},
\end{align}
where $\Tilde{z} \equiv z - i  \omega_0 / \hbar $ and we used the general massive dispersion relation~(\ref{Eq:massive_dispersion}). 
Integrating this expression with respect to $p$ is much easier than dealing directly with the memory kernel~(\ref{Eq:Memory_kernel}). We indeed obtain
\begin{align}
    \int_{-\Lambda}^\Lambda dp ~\frac{\Tilde{z} \hbar^2}{\Tilde{z}^2 \hbar^2 + m^2 + p^2} = \frac{2 \hbar^2 \Tilde{z} \text{arctan}\left( \frac{\Lambda}{\sqrt{m^2 + \Tilde{z}^2 \hbar^2}}\right)}{\sqrt{m^2 + \Tilde{z}^2 \hbar^2}}.
\end{align}
Then, the Laplace-transformed dynamical equation~(\ref{Eq: Laplace transformed master}) becomes
\begin{align}
     z \overline{\Phi} (z) - \Phi (0) = -4g^2 \frac{\Tilde{z} \text{arctan}\left( \frac{\Lambda}{\sqrt{m^2 + \Tilde{z}^2 \hbar^2}}\right)}{\sqrt{m^2 + \Tilde{z}^2 \hbar^2}} \overline{\Phi} (z).
\end{align}
Rearranging the above equation yields
\begin{align}\label{Appdx: Resolvent definition}
    \overline{\Phi} (z) = \left[z + 4g^2 \frac{\Tilde{z} \text{arctan}\left( \frac{\Lambda}{\sqrt{m^2 + \Tilde{z}^2 \hbar^2}}\right)}{\sqrt{m^2 + \Tilde{z}^2 \hbar^2}}\right]^{-1} \Phi (0).
\end{align}

The final solution is obtained via the inverse Laplace transform:
\begin{align}
    \Phi (t) = \text{Lap}^{-1}\left[\overline{\Phi} (z)\right] (t).
\end{align}
Once again, utilizing the Laplace shifting relation $\text{Lap}^{-1} \left[f(z-a)\right](t) = e^{at}\text{Lap}^{-1} \left[f(z)\right](t)$, we find
\begin{align}
    \Phi (t) = e^{\frac{i}{\hbar}\omega_0 t } \text{Lap}^{-1}\left[\overline{\Phi} \left(z + \frac{i}{\hbar} \omega_0 \right)\right] (t),
\end{align}
where the function inside the inverse Laplace transform is
\begin{align}\label{Eq: Before inverse Laplace}
    \overline{\Phi} \left(z + \frac{i}{\hbar} \omega_0 \right) = \left[z + i \frac{\omega_0}{\hbar} + 4g^2 \frac{z \text{arctan}\left( \frac{\Lambda}{\sqrt{m^2 + z^2 \hbar^2}}\right)}{\sqrt{m^2 + z^2 \hbar^2}} \right]^{-1} \Phi (0).
\end{align}
The inverse Laplace transform is evaluated via the Bromwich integral, leading to the formal solution of the dynamical equation~(\ref{Eq:master_equation}):
\begin{align}
    \Phi(t) &= e^{\frac{i}{\hbar}\omega_0 t} \lim_{R \rightarrow \infty} \int_{-i R + \sigma}^{i R + \sigma} \frac{dz}{2\pi i} e^{z t} \left[z + i \frac{\omega_0}{\hbar} + 4g^2 \frac{z \text{arctan}\left( \frac{\Lambda}{\sqrt{m^2 + z^2 \hbar^2}} \right)}{\sqrt{m^2 + z^2 \hbar^2}} \right]^{-1} \Phi(0) \nonumber \\
    &= e^{\frac{i}{\hbar}\omega_0 t} \lim_{R \rightarrow \infty} \int_{-i R + \sigma}^{i R + \sigma} \frac{dz}{2\pi i} e^{z \frac{t}{\hbar}} \left[z + i \omega_0 + 4g^2 \frac{z \text{arctan}\left( \frac{\Lambda}{\sqrt{m^2 + z^2}} \right)}{\sqrt{m^2 + z^2}} \right]^{-1} \Phi(0),
\end{align}
where, in the last equality, we have rescaled the integration variable by $\hbar$. Rescaling the integration variable once more by $z\rightarrow g^2 z$, we arrive at the expression given in Eqs.~(\ref{Eq:full_solution}) and~(\ref{Eq:resolvent}).

\section{Position representation of the wave function}\label{Appdx: wave function}

In this section, we evaluate the wave function~(\ref{Eq: Wave function}) in the position representation. Acting $\bra{\mathrm{g}x}$, which is defined in Eq.~(\ref{Eq. position representation}), on the Ansatz~(\ref{Eq: Ansatz}), we find the following expression:
\begin{align}
    \braket{\mathrm{g}~x}{ \phi(t)} = e^{- \frac{i}{\hbar} \omega_0 t} \braket{\mathrm{g}~x}{\Phi(t)} = \int_{-\Lambda}^{\Lambda} dp~ \left(e^{-\frac{i}{\hbar}(\omega_p t + px)}\Phi_p (t) + e^{\frac{i}{\hbar}(\omega_p t + px)}\Psi_p (t)\right).
\end{align}
The functions $\Phi_p (t)$ and $\Psi_p(t)$ can be written in terms of $\Phi(t)$, as given in Eq.~(\ref{Eq. solution to phi p}). The above expression then becomes
\begin{align}
    \braket{\mathrm{g}~x}{ \phi(t)} = \frac{2g}{i \hbar} \int_0^t ds~ e^{- \frac{i}{\hbar}\omega_0 s} \Phi(s) \int_{-\Lambda}^{\Lambda} dp~ \cos \left(\frac{\omega_p (t-s)+ px}{\hbar}\right).
\end{align}

Notice that the above expression is a convolution in time. We thereby apply the Laplace transform~(\ref{Eq: defintion of Laplace transform}):
\begin{align}
    \text{Lap}\left[\braket{\mathrm{g}~x}{ \phi(t)} \right] (z) &= \frac{2g}{i\hbar} \text{Lap}\left[e^{- \frac{i}{\hbar}\omega_0 s} \Phi(t)\right](z) \int_{-\Lambda}^{\Lambda} dp~ \text{Lap}\left[ \cos \left(\frac{\omega_p t + px}{\hbar} \right) \right](z) \\
    &= \frac{2g}{i\hbar} \overline{\Phi}\left(z + \frac{i \omega_0}{\hbar}\right) \int_{-\Lambda}^{\Lambda} dp~ \hbar \frac{\hbar z \cos\left(\frac{p x}{\hbar} \right) - \omega_p \sin\left(\frac{p x}{\hbar} \right)}{z^2 \hbar^2 + \omega_p^2},
\end{align}
where $\overline{\Phi}(z)$ is given in Eq.~(\ref{Appdx: Resolvent definition}).
Using the inverse Laplace transform, we find the formal solution for the wave function:
\begin{align}
    \braket{\mathrm{g}~x}{ \phi(t)} &= \lim_{R \rightarrow \infty} \int_{-i R + \sigma}^{i R + \sigma} \frac{dz}{2\pi i} e^{z t} \left[ \frac{2g}{i} \overline{\Phi}\left(z + \frac{i \omega_0}{\hbar}\right) \int_{-\Lambda}^{\Lambda} dp~ \frac{\hbar z \cos\left(\frac{p x}{\hbar} \right) - \omega_p \sin\left(\frac{p x}{\hbar} \right)}{z^2 \hbar^2 + \omega_p^2} \right] \\
    &= \frac{2g}{i \hbar^2} \lim_{R \rightarrow \infty} \int_{-i R + \sigma}^{i R + \sigma} \frac{dz}{2\pi i} e^{ \frac{z t}{\hbar}} F(z) \int_{-\Lambda}^{\Lambda} dp~ \frac{z \cos\left(\frac{p x}{\hbar} \right) - \omega_p \sin\left(\frac{p x}{\hbar} \right)}{z^2 + \omega_p^2}\\
    &= \frac{2g}{i \hbar^2} \lim_{R \rightarrow \infty} \int_{-i R + \sigma}^{i R + \sigma} \frac{dz}{2\pi i} e^{ \frac{z t}{\hbar}} F(z) \int_{-\Lambda}^{\Lambda} dp~ \frac{z \cos\left(\frac{p x}{\hbar} \right)}{z^2 + \omega_p^2},
\end{align}
where in the last equality, we used the fact that the sine function is an odd function of $x$ and $p$. Since it is integrated over $p\in \left[-\Lambda , \Lambda\right]$, the sine term vanishes. Finally, we normalize the wave function by replacing the constant $2g/i\hbar^2 $ with the normalization constant $-i \sqrt{2g^2 \pi / \hbar}$,  leading to the expression in Eq.~(\ref{Eq: Wave function}).
\section{Markovianity and the semigroup property}\label{Appdx: CP-divisibility}
    In this appendix, we introduce the notion of one-parameter quantum dynamical semigroup property (semigroup property) for the system dynamics. Using the explicit solutions for the probability amplitudes in Eqs.~(\ref{Eq: Surv. Prob. 1}), (\ref{Eq: Asym-exact 1}), and (\ref{Eq: Asym-exact 2}), we demonstrate that the semigroup property is satisfied only for the parameter limit of $m\rightarrow 0$ and $\Lambda \rightarrow \infty$. This limit corresponds to the Markovian dynamics introduced in Sec.~\ref{sec: L infinite and m zero}. 

    Let us state the definition of the semigroup property. Consider a linear map between the initial and the final state $\rho(t) = \mathcal{E}_t \left[\rho(0)\right]$, where the linear map $\mathcal{E}_t$, is called a quantum channel. A physical map between any two quantum states is required to be completely positive and trace-preserving (CPTP). By the Choi theorem \cite{Choi1975}, any CPTP map admits a non-unique Kraus representation, 
    \begin{align}
        \mathcal{E}_t \left[\rho(0)\right] = \sum_{i} K_i(t) \rho(0) K_i^\dagger (t), \quad \sum_i K^\dagger_i (t) K_i (t) = \mathbb{I},\label{Eq: Kraus representation definition}
    \end{align}
    where $\{K_i (t)\}$ is a set of Kraus operators satisfying the above completeness relation. 
    
    An explicit Kraus representation can be obtained for our setup. To this end, consider an initially superimposed qubit state coupled to the environmental vacuum,
    \begin{align}
        \ket{\phi(0)} = \Big(\phi_0 \ket{\mathrm{g} }_S+\phi (0) \ket{\mathrm{e} }_S\Big) \otimes \ket{\text{vac}}_E .
    \end{align}
    where $\ket{\mathrm{e}}_S, \ket{\mathrm{g}}_S$ and $\ket{\text{vac}}_E$ are defined in Sec.~\ref{Sec:model}. The particle number of this initial state is evaluated with the number operator $N = \sigma_+ \sigma_- + \int_{-\infty}^{\infty} dp \left(b^\dagger_p b_p + c_p c^\dagger_p\right)$, yielding 
    \begin{align}
        \bra{\phi(0)} N \ket{\phi(0)} = 1- \left|\phi_0\right|^2.
    \end{align}
    In the main text of the present manuscript, we set $\phi_0 =0$ and $\phi (0) =1$ (the single-particle sector), since our focus was on the survival probability of a single excitation. Allowing $\phi_0 \not=0$, a general state with particle number $1- \left|\phi_0\right|^2$ can be written as 
    \begin{align}\label{Eq: general entangled Ansatz}
        \ket{\phi(t)} = \phi_0 \ket{\mathrm{g}~ \text{vac}} + \phi (t) \ket{\mathrm{e} ~\text{vac}} + \int_{-\Lambda}^{\Lambda} dp \phi_p (t) b^\dagger_p \ket{\mathrm{g} ~\text{vac}} + \int_{-\Lambda}^{\Lambda} dp \psi_p (t) c_p \ket{\mathrm{g} ~\text{vac}}, 
    \end{align}
    where one can readily verify that $\bra{\phi(t)}N \ket{\phi(t)} = 1- \left|\phi_0\right|^2$. 

    Starting from the Ansatz~(\ref{Eq: general entangled Ansatz}), we can obtain the reduced density matrix of the system. Let $\rho(t) = \ketbra{\psi(t)}{\psi(t)}$ denote the joint system-environment state. Tracing out the environment gives
    \begin{align}
        \rho_S(t) &= \mathrm{Tr}_E[\rho(t)] = \prescript{}{E}{\bra{\text{vac}}}  \rho (t)  \ket{\text{vac}}_E + \int_{-\Lambda}^{\Lambda} dp \prescript{}{E}{\bra{\text{vac}}}b_p \rho b^\dagger_p \ket{\text{vac}}_E+ \int_{-\Lambda}^{\Lambda} dp \prescript{}{E}{\bra{\text{vac}}}c^\dagger_p \rho c_p \ket{\text{vac}}_E \\
        &= \begin{pmatrix}
    1 - |\phi (t)|^2 & \phi_0\phi^*(t) \\
    \phi_0^*\phi(t) & |\phi (t)|^2
    \end{pmatrix},\label{Eq: General Ansatz density matrix}
    \end{align}
    where we have used the basis $\ket{\mathrm{g}}_S = (1,0)^T$, $\ket{\mathrm{e}}_S=(0,1)^T$.

    For the above system density matrix~(\ref{Eq: General Ansatz density matrix}), the Kraus operator that connects the initial state $\rho_S (0)$ and the final state $\rho_S (t)$ can be given by the following two operators:
    \begin{align}
        K_0 (t)=\left(\begin{array}{cc}
            1 &0  \\
             0& \phi(t) / \phi(0)
        \end{array}\right) , \quad K_1 (t)= \left(\begin{array}{cc}
            0 &\sqrt{1-\left|\phi(t) / \phi(0)\right|^2}  \\
             0& 0
        \end{array}\right) .
    \end{align}
    Inserting them to Eq.~(\ref{Eq: Kraus representation definition}), one can show that $\mathcal{E}_t\left[\rho_S(0)\right] = \rho_S (t)$. Accordingly, for our model and initial condition, the quantum channel $\mathcal{E}_t$ is CPTP for all time $t$.  

    The semigroup property is a condition on the quantum channel $\mathcal{E}_t$, requiring that for any intermediate time $0<s<t$, the evolution can be factorized as
    \begin{align}
        \mathcal{E}_{t-s}\circ \mathcal{E}_{s} = \mathcal{E}_t.
    \end{align}
    Here, $\mathcal E_{t-s}$ and $\mathcal E_s$ are members of the same one-parameter family of the CPTP maps. For our model, this factorization condition can be translated into a constraint on the survival amplitude of the system as
    \begin{align}\label{Eq: semi-group}
        \phi(t-s) \phi(s) = \phi(t).
    \end{align}
    Therefore, to establish the semigroup property in our setting, it is sufficient to verify that $\phi(t)$ satisfies the above composition law for all $t\geq 0$. We examine the composition law~(\ref{Eq: semi-group}) in the following three parameter regimes: (i) $m=0, \Lambda\rightarrow \infty$; (ii) $m\not= 0, \Lambda\rightarrow \infty$; and (iii) $m=0, \Lambda< \infty$. 
    
    (i) For $m=0$ and in the limit $ \Lambda\rightarrow \infty$ discussed in Sec.~\ref{sec: L infinite and m zero}, the exact solution for the probability amplitude was obtained in Eq.~(\ref{Eq: Surv. Prob. 1}). Due to its purely exponential form, this solution satisfies the composition law~(\ref{Eq: semi-group}) trivially. Consequently, in this parameter regime, the dynamics of our model satisfy the semigroup property and are Markovian. 

    (ii) For $m\not= 0$ and in the limit $ \Lambda\rightarrow \infty$, which we discuss in Sec.~\ref{Sec:Low_energy_structure_and_Markovianity}, explicit forms of $\phi(t)$ in the short and long-time regimes are given in Eq.~(\ref{Eq: Asym-exact short 1}) and (\ref{Eq: Asym-exact long 1}), respectively. At the leading order in the short-time regime, Eq.~(\ref{Eq: Asym-exact short 1}) has an exponential form and is therefore consistent with the composition law~(\ref{Eq: semi-group}). However, in the long-time regime~(\ref{Eq: Asym-exact long 1}), the presence of a $t^{-3/2}$ term prevents the composition. As a result, the dynamics in this parameter regime do not satisfy the semigroup property and therefore are non-Markovian. 

    (iii) Finally, for $m=0$ and $ \Lambda< \infty$, which we discuss in Sec.~\ref{Sec:High_energy_structure_and_Markovianity}, the short and long-time asymptotic forms of $\phi(t)$ are given in Eq.~(\ref{Eq: Asym-exact short 2}) and (\ref{Eq: Asym-exact long 2}). In this case, the short-time behavior already violates the composition law~(\ref{Eq: semi-group}) due to the $t^2$ term contribution. Similarly, the long-time asymptote does not satisfy the composition law either. Therefore, in this parameter regime, the system dynamics do not satisfy the semigroup property and consequently are non-Markovian.

\section{Survival probability of GKSL master equation}\label{Appdx: Markov}
In this appendix, we prove the statement made in Sec.~\ref{Sec: Connection to the Markovianity}: under Markovian (GKSL) dynamics, the survival probability of the excited state of a qubit decays exponentially.

Consider a qubit density matrix $
\rho_S (t)$ with the initial condition 
\begin{align}\label{Eq: Appendix GKSL density matrix}
    \rho_S (0) = \left(\begin{array}{cc}
         0& 0 \\
        0& 1
    \end{array}\right),
\end{align}
written in the basis $\left\{\ket{g}_S =(1,0)^T, \ket{e}_S = (0,1)^T\right\}$. The survival probability of the excited state at time $t$ is then given by the Born rule:
\begin{align}
P(t) &= \mathrm{Tr}_S\Big[\ket{e}_{S}\prescript{}{S}{\bra{e}}\rho_S(t)\Big],
\end{align}
where $\ket{e}_{S}\prescript{}{S}{\bra{e}}$ is the projection operator. We show that if $\rho_S (t)$ evolves according to the GKSL master equation $\partial_t \rho_S = \mathcal{L}\rho_S (t)$ with the Liouvillian superoperator $\mathcal{L}$, then the survival probability $P(t)$ exhibits an exponential decay.  

Note that the space of complex \(n \times n\) matrices, equipped with the Hilbert-Schmidt inner product, can be mapped to \(\mathbb{C}^{n^2}\) via the vectorization operation:
\begin{align}
\mathrm{Vec}[A] \equiv \vec{A} := \sum_{i,j=1}^n A_{ij} \ket{i}\otimes\ket{j} = \sum_{i,j=1}^n A_{ij}\ket{ij}, \quad \mathrm{Tr}(B^\dagger A) = \vec{B}^\dagger \vec{A}.
\end{align}
Defining the basis vectors \(\vec{E}_{ij} := \mathrm{Vec}[\ket{i}_S\prescript{}{S}{\bra{j}}]\), we can express the initial density operator in Eq.~(\ref{Eq: Appendix GKSL density matrix}) as \(\vec{\rho}_S (0)=\vec{E}_{11} =(0,0,0,1)^T\). The dynamical evolution \(\rho_S(t) = e^{\mathcal{L}t}\rho_S(0)\) can be vectorized as \(\vec{\rho}_S(t) = e^{\bar{\bar{\mathcal{L}}}t}\vec{\rho}_S(0)\),
where \(\bar{\bar{\mathcal{L}}}\) is the matrix representation of the Liouvillian superoperator, a generally non-Hermitian \(4 \times 4\) matrix.
Thus, the survival probability evolves as
\begin{align}
P(t) &= \mathrm{Tr}_S\left[\ket{e}_{S}\prescript{}{S}{\bra{e}} \rho_S(t)\right] = \vec{E}_{11}^\dagger e^{\bar{\bar{\mathcal{L}}}t}\vec{\rho}_S(0) = \vec{E}_{11}^\dagger e^{\bar{\bar{\mathcal{L}}}t}\vec{E}_{11},
\end{align}
where in the last line we have used \(\vec{\rho}_S (0) = \vec{E}_{11}\).

The eigenvalues of the GKSL Liouvillian \(\bar{\bar{\mathcal{L}}}\) satisfy two key properties: (i) the real parts are non-positive, and (ii) complex eigenvalues appear in conjugate pairs~\cite{minganti2018spectral}. 
Note that \(\bar{\bar{\mathcal{L}}}\) is non-Hermitian, and hence it may be non-diagonalizable, namely at an exceptional point where its rank is reduced.

The matrix \(\bar{\bar{\mathcal{L}}}\) can be diagonalized or put into a Jordan normal form using a biorthonormal basis or Jordan basis:
\begin{align}
\bar{\bar{\mathcal{L}}}\vec{R}_i = \lambda_i \vec{R}_i, \quad \vec{L}_i^\dagger \bar{\bar{\mathcal{L}}} = \lambda_i^* \vec{L}_i^\dagger, \quad \vec{L}_i^\dagger \vec{R}_j = \delta_{ij}.
\end{align}
Assembling \(\{\vec{R}_i\}\) as columns of \(R\) and \(\{\vec{L}_i^\dagger\}\) as rows of \(R^{-1}\), we write \(\bar{\bar{\mathcal{L}}} = R D R^{-1}\), where \(D\) is either diagonal or in a Jordan normal form. 

For example, let us consider the following two cases:
\begin{align}
D_1 &= \begin{pmatrix}
\lambda_1 & 0 & 0 & 0 \\
0 & \lambda_2 & 0 & 0 \\
0 & 0 & \lambda_3 & 0 \\
0 & 0 & 0 & \lambda_4
\end{pmatrix}, \quad
D_2 = \begin{pmatrix}
\lambda_1 & 1 & 0 & 0 \\
0 & \lambda_1 & 0 & 0 \\
0 & 0 & \lambda_3 & 0 \\
0 & 0 & 0 & \lambda_4
\end{pmatrix}.
\end{align}
Exponentiating yields \( e^{\bar{\bar{\mathcal{L}}}t} = R e^{D_\alpha t} R^{-1}, \quad \alpha=1,2, \) with
\begin{align}
e^{D_1 t} &= \begin{pmatrix}
e^{\lambda_1 t} & 0 & 0 & 0 \\
0 & e^{\lambda_2 t} & 0 & 0 \\
0 & 0 & e^{\lambda_3 t} & 0 \\
0 & 0 & 0 & e^{\lambda_4 t}
\end{pmatrix}, \quad
e^{D_2 t} = \begin{pmatrix}
e^{\lambda_1 t} &  t e^{\lambda_1 t} & 0 & 0 \\
0 & e^{\lambda_1 t} & 0 & 0 \\
0 & 0 & e^{\lambda_3 t} & 0 \\
0 & 0 & 0 & e^{\lambda_4 t}
\end{pmatrix}.
\end{align}
In the diagonal case $D_1$, the survival probability takes the following form:
\begin{align}
P_{\text{diag}}(t) = \sum_{i=1}^4 \left(\vec{E}_{11}^\dagger R^{-1}\right)_i  \left( R \vec{E}_{11}\right)_i  e^{\lambda_i t } = \sum_{i=1}^{4} \left(\vec{L}_4^\dagger \right)_{i} \left(\vec{R}_4 \right)_i e^{\lambda_i t} = \sum_{i=1}^4 c_i e^{\lambda_i t},
\end{align}
where \(c_i := \left(\vec{L}_4^\dagger \right)_{i} \left(\vec{R}_4 \right)_i\). We have also used the fact that \(\vec{E}_{11}^\dagger = (0,0,0,1)^T\) and that the matrices \(R^{-1}\) and \(R\) have vectors \(\vec{L}_i^\dagger\) and \(\vec{R}_i\) as rows and columns. From this expression, it is easy to verify that when \(t=0\), the above expression reduces to unity by the relation \(\vec{L}_4^\dagger \vec{R}_4 = 1\), consistent with the initial condition. In the non-diagonal case $D_2$, the survival probability takes the following form:
\begin{align}
P_{\text{non-diag}}(t) &= \sum_{i=1}^{4} \left(\vec{E}_{11}^\dagger R^{-1}\right)_i \left(e^{D_2 t}\right)_{ij}  \left( R \vec{E}_{11}\right)_j \\
&= \sum_{i=1}^{4} \left(\vec{L}_4^\dagger \right)_{i} \left(e^{D_2 t}\right)_{ij}\left(\vec{R}_4 \right)_j \\
&= \left(c_{11}+c_{22}+c_{12} t \right)e^{\lambda_1 t} + c_{33} e^{\lambda_3 t} + c_{44} e^{\lambda_4 t},
\end{align}
where \(c_{ij} := \left(\vec{L}_4^\dagger \right)_{i} \left(\vec{R}_4 \right)_j\).

Finally, the spectral properties of the GKSL Liouvillian imply that the eigenvalues can either be purely decaying terms \(\lambda_i\) or oscillatory parts \(\lambda_i^{\pm}\),
\begin{align}
\lambda_i &= - a_i , \quad a_i \geq 0, \\
\lambda_i^{(\pm)} &= -\alpha_i \pm i \beta_i , \quad \alpha_i , \beta_i \geq 0.
\end{align}
In the diagonal case $D_1$, let us consider the following eigenvalues as an example: \(\lambda_1 = - \alpha_1\), \(\lambda_2 = -\alpha_2\), and \(\lambda_3 = \lambda_4^* = -\alpha + i \beta\). Then we have
\begin{align}
P_{\text{diag}}(t) = c_1 e^{-\alpha_1 t } + c_2 e^{-\alpha_2 t} + e^{-\alpha t } \left(c_3 e^{i \beta t } + c_4 e^{-i \beta t }\right),
\end{align}
where the last term corresponds to exponential decay with oscillation. 
In the non-diagonal case $D_2$, since \(\lambda_1\) is the eigenvalue at the exceptional point where two complex conjugate pairs meet, it can only be a negative real number \(\lambda_1 = -a_1\). For the other two eigenvalues, we can assume \(\lambda_3 = \lambda_4^* = -\alpha + i \beta\), and therefore the survival probability is given by
\begin{align}
P_{\text{non-diag}}(t) = \left(c_{11}+c_{22}+c_{12} t \right)e^{-\alpha_1 t} + \left(c_{33} e^{i \beta t} + c_{44} e^{- i \beta t}\right) e^{-\alpha t}.
\end{align}
Here, the first term contains a polynomial prefactor. The degree of the polynomial depends on the size of the Jordan block formed in the Liouvillian superoperator.

In both the diagonalizable and non-diagonalizable cases, we observe that the decaying profile of the survival probability under Markovian dynamics is exponential; power laws are non-existent.

\section{Contour deformation}\label{Appdx: Contour}

This appendix shows how the two contours in Fig.~\ref{fig:Poles} are equivalent. We begin by adding a long contour shown as a dashed line in Fig.~\ref{fig: Contour deformation}(a). 
This long contour does not contribute to the overall integral (\ref{Eq:Simpler_solution}) because of the exponential factor in the integrand, $e^{z t / \hbar}$, which exponentially suppresses the additional contour.

Next, we add extra contours around the branch cuts, shown as solid lines in Fig.~\ref{fig: Contour deformation}(b). 
Because of these extra contours, the overall integral (\ref{Eq:Simpler_solution}) now includes an additional contribution:
\begin{align}
    \int_{-i R + \sigma}^{i R+ \sigma}\dots + \int_{\text{BC}}^{}\dots ~.
\end{align}
Since the contour shown in Fig.~\ref{fig: Contour deformation}(b) is closed, we can apply Cauchy's theorem to deform the contour to a closed path around the pole, as shown in Fig.~\ref{fig: Contour deformation}(d).

Finally, we subtract the extra contribution from the branch cuts by adding contours around the branch cuts, shown in Fig.~\ref{fig: Contour deformation}(c), which are oriented in the opposite direction to the one added in Fig.~\ref{fig: Contour deformation}(b). 
The resulting contour, shown again in Fig.~\ref{fig: Contour deformation}(b), is our final result in Fig.~(\ref{fig:Poles}).

\begin{figure}[ht]
    \centering
    \includegraphics[width=0.7\textwidth]{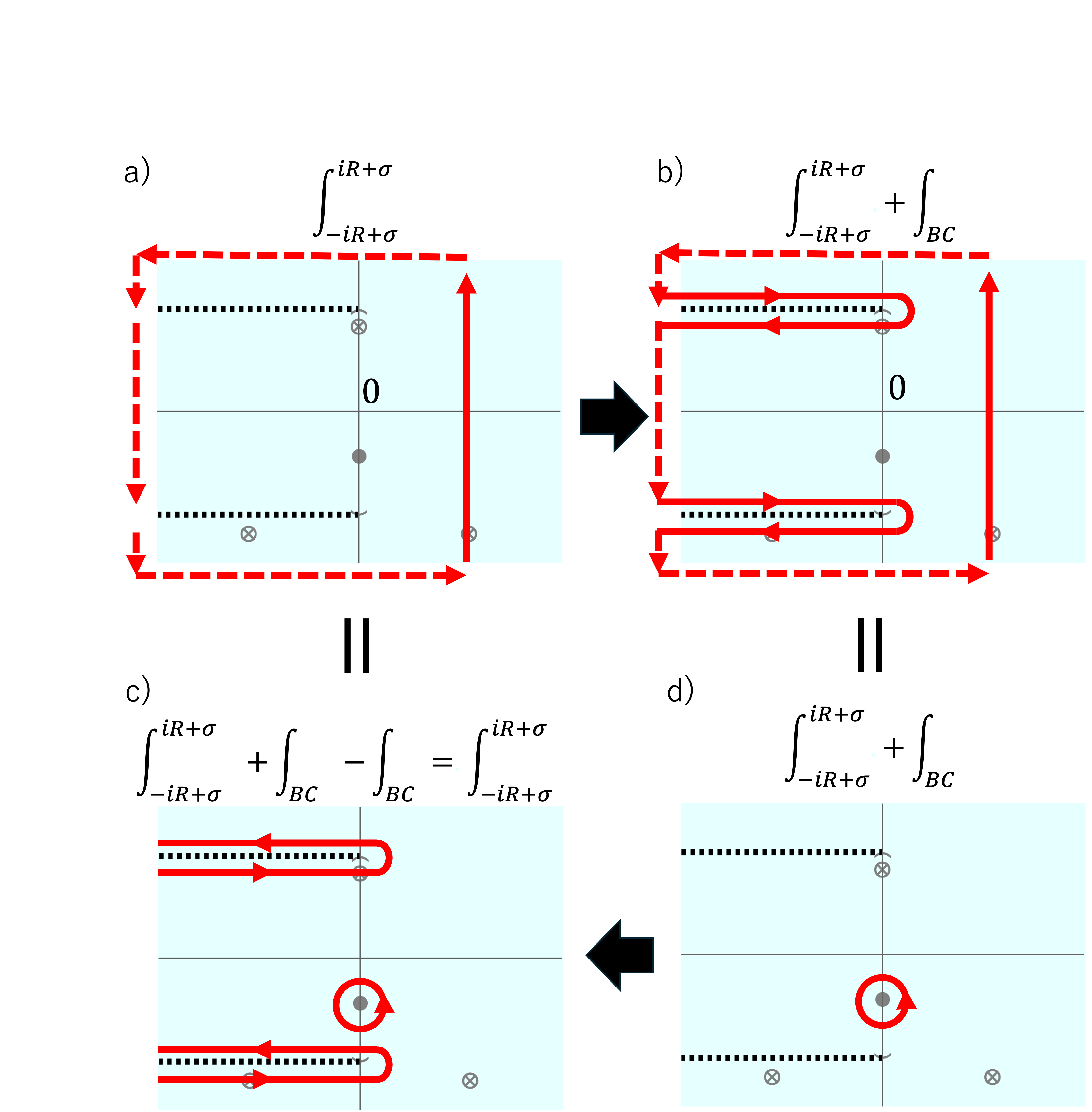}
    \caption{Process of contour deformation demonstrating the equivalence of two contours in Fig.~\ref{fig:Poles}. For a detailed explanation, see Appendix~\ref{Appdx: Contour}.}
    \label{fig: Contour deformation}
\end{figure}

\section{Quantum Zeno Effect}\label{Appdx: Quantum Zeno Effect}

The quantum Zeno effect, which we mention in Secs.~\ref{Sec: Bound state} and ~\ref{Sec: short time with finite cut-off}, is a seemingly paradoxical phenomenon in quantum mechanics. It was first theoretically proposed by Misra and Sudarshan in 1977~\cite{misra1977zeno} and demonstrated experimentally by Itano, Bollinger, and Wineland in 1990~\cite{Itano1990}. 
It states that a decaying quantum state under continuous observation leads to a non-decaying probability over time. 

To elucidate, consider a Hamiltonian $H$ with a bounded spectrum and an unstable state $\ket{M}$ as the initial state. 
Define the survival probability as $Q(t):= |\bra{M}e^{-i H t / \hbar}\ket{M}|^2$. 
Assuming that the Hamiltonian's expectation value with respect to the initial state $\ket{M}$ is finite, \textit{i.e.}, $\bra{M}H\ket{M}<\infty$ and $\bra{M}H^2 \ket{M}< \infty$, it can be shown that $dQ(t)/dt \big|_{t=0} = 0$.

By segmenting the time interval into $n$ short time slices and performing a projective measurement at each slice, the probability that the result of every measurement is the excited state is given by $Q(t/n)^n$. In the limit of continuous measurement, as $n$ approaches infinity, we find
\begin{equation}
    \lim_{n\rightarrow \infty} Q\left[\frac{t}{n}\right]^n = \lim_{n\rightarrow \infty} \left(1+ \frac{1}{2}\frac{d^2 Q(0)}{dt^2}\left(\frac{t}{n}\right)^2\right)^n =1.
\end{equation}

This suggests that continuous observation ensures that the initial state $\ket{M}$ remains undecayed, as the survival probability approaches unity. 

\end{widetext}
\bibliographystyle{apsrev4-2}
\bibliography{references}
\end{document}